\documentclass[11pt,letterpaper]{article}

\usepackage{amsmath,amssymb,amsfonts,amsthm}
\IfFileExists{bbm.sty}{\usepackage{bbm}}{}
\usepackage{bm}
\usepackage{booktabs}
\usepackage{array}
\usepackage{graphicx}
\usepackage{float}
\usepackage{placeins}
\usepackage[round,authoryear]{natbib}
\usepackage[letterpaper,margin=1in]{geometry}
\usepackage{microtype}
\usepackage[skip=8pt]{caption}
\usepackage{orcidlink}
\usepackage{xcolor}
\usepackage{hyperref}
\hypersetup{hidelinks,hypertexnames=false}

\newcommand{\E}{\mathbb{E}}
\newcommand{\F}{\mathcal{F}}
\newcommand{\D}{\mathrm{d}}
\newcommand{\tabnote}[1]{%
  \par\vspace{0.55em}\begin{minipage}{0.96\textwidth}%
  \raggedright\footnotesize\noindent\textit{Note:} #1%
  \end{minipage}}

\graphicspath{{figures/}}

\newcommand{\figACFCaption}{Empirical and model-implied volatility persistence at
$p=1.2$. Each panel compares the empirical autocorrelation function of squared returns
with medians and 5th--95th percentile simulation envelopes from the fitted ALM-GARCH
and GARCH$(1,1)$ models over lags to $250$ days. The LM-GARCH median is also shown; its
envelope is omitted for legibility. Simulations use the empirical sample length after a
$5{,}000$-observation burn-in.}

\newcommand{\figResponseCaption}{Fitted sign-specific volatility responses at
$p=1.2$. For each asset and sign,
$g_s(k)=\xi_s\{\gamma_s/(\gamma_s+k\tau)\}^{p}$ is plotted against $k+1$ on logarithmic
axes. The initial vertical difference reflects the level channel; differences in
finite-horizon persistence reflect the memory channel. Both branches share
the common asymptotic exponent $p$. The fitted FTSE~100 positive-branch amplitude,
$\hat\xi_+=4.6\times10^{-7}$, lies below the common vertical scale and is therefore not
visible.}

\theoremstyle{plain}
\newtheorem{theorem}{Theorem}
\newtheorem{lemma}{Lemma}
\newtheorem{proposition}{Proposition}
\newtheorem{condition}{Condition}
\newtheorem{assumption}{Assumption}

\begin{document}
\pagenumbering{arabic}

\begin{center}
{\Large\bfseries Asymmetric Long-Memory GARCH:\\[0.3em]
Sign-Dependent Kernel Injection in a Two-Dimensional Markov Chain\par}
\vspace{1em}
{\large Kennedy Titus Kayaki\orcidlink{0009-0009-5132-6338}$^{*}$\quad
Kyungsub Lee\orcidlink{0000-0001-9499-0671}$^{*}$\par}
\vspace{0.5em}
$^{*}$Department of Statistics, Yeungnam University, Gyeongsan, Republic of Korea\\
\end{center}

\begin{abstract}
We introduce ALM-GARCH, an asymmetric long-memory GARCH model in which positive and negative innovations enter conditional variance with different injection amplitudes and kernel offsets. These departures define testable level and memory channels relative to a nested symmetric benchmark. Positive Harris recurrence holds for interior configurations under a Foster--Lyapunov condition. Across five equity indices and Bitcoin, joint symmetry is rejected throughout, driven primarily by the level channel. The memory channel is supported for the Nikkei~225, KOSPI, and Bitcoin but is weakly identified when the positive branch is nearly inactive. Out-of-sample performance is broadly comparable to standard benchmarks.
\end{abstract}



\medskip
\noindent\textit{Supplementary Appendix:} Detailed proofs, full parameter estimates,
simulation and sensitivity results, optimization diagnostics, and additional figures
follow main article.

\bigskip

\section*{Introduction}\addcontentsline{toc}{section}{Introduction}
\label{sec:intro}

The volatility of financial returns is both persistent and asymmetric. Its
autocorrelations decay slowly, at a hyperbolic rather than geometric rate, and its
response to news is sign-dependent: negative returns raise future volatility by more
than positive returns of equal size. Capturing both features in a single tractable
model has proved difficult. Long-memory specifications such as FIGARCH
\citep{baillie1996} reproduce the slow decay but require an infinite-order
representation of the past, while the standard asymmetric models, GJR-GARCH
\citep{glosten1993} and EGARCH \citep{nelson1991}, capture sign-dependence but only
within a short-memory recursion.

\citet{lee2026} recently showed that long memory can be obtained from a
finite-dimensional object: a two-dimensional Markov chain in which a second state
variable acts as a memory carrier, compressing the past into an effective kernel age.
The construction reproduces hyperbolic decay without carrying the full history. However
, that model is symmetric and so is silent on the leverage effect that is among the
most robust empirical regularities in volatility.

This paper therefore develops the asymmetric extension. We allow the power-law kernel injected by
each shock to depend on the sign of that shock, in both its amplitude and its offset.
The result, which we call asymmetric long-memory GARCH (ALM-GARCH), retains the
finite-dimensional Markov structure of the base model and nests it exactly, while
introducing two conceptually distinct forms of asymmetry. The first form is a
\emph{level (amplitude) channel}: negative shocks may load more heavily on future variance, the
leverage effect. The second is a \emph{memory (offset) channel}: shocks of different signs may reset the
volatility kernel to different ages, producing different finite-horizon persistence
profiles. Because the power exponent $p$ is common across signs, ``memory asymmetry''
here refers to sign dependence in kernel age and finite-horizon persistence, not to
different asymptotic power-law exponents. This mechanism differs from established
asymmetric-power and component approaches. Asymmetric-power models act on response
amplitudes, while component models introduce multiple timescales without a sign-specific
reset of a finite-dimensional kernel-age state.

The theoretical analysis derives the asymmetric recursion from sign-dependent kernel
injection, extends the drift and regularity arguments of \citet{lee2026} to the interior
of the resulting two-regime parameter space, and characterizes how the level and memory
channels move a diagnostic frontier associated with the sufficient stability
certificate under the stated conditions. The
symmetric results are recovered at the null. Identification analysis develops nested
likelihood-ratio tests for the two channels. Their scores are highly correlated near the
symmetric null, so we use the joint test as the main test of asymmetry and interpret the
channel-specific tests more cautiously.

We estimate the model on five international equity indices and Bitcoin. For each of the six series we reject the joint null
that both the injection amplitudes and kernel offsets are symmetric. This overall result is carried primarily by the level channel and does
not imply that each of the two restrictions is rejected individually for every series. The
finite-sample bootstrap rejects the memory restriction for Nikkei~225, KOSPI, and
Bitcoin, but not for DAX at the five-percent level. For the S\&P~500 and FTSE~100,
the nearly inactive positive branch prevents separate identification of the memory
channel. We report the cross-market ordering as a descriptive feature of the sample,
not as evidence of regional market differences.

A direct in-sample comparison with FIAPARCH, an established asymmetric long-memory
specification, favors FIAPARCH by BIC in all six series. We acknowledge ALM-GARCH's empirical
role as narrower: it provides a finite-dimensional and separately testable decomposition
of asymmetry rather than a uniformly best-fitting long-memory specification.

Out of sample, ALM-GARCH is broadly comparable with the return-based benchmarks. It is
significantly more accurate than GARCH$(1,1)$ and GJR-GARCH on the DAX and than
GJR-GARCH on KOSPI, whereas EGARCH is significantly more accurate on the S\&P~500 and
the Nikkei and FIGARCH is more accurate on KOSPI and Bitcoin. The remaining differences
are not significant. HAR-RV, which forecasts realized variance directly from its own
lags, records the lowest loss in every series. The asymmetric and symmetric long-memory
specifications also produce similar forecasts, with a significant improvement from the
extension only on DAX. ALM-GARCH therefore serves mainly as a structural
decomposition of asymmetry rather than as a general improvement in fit or forecast
accuracy.

\subsection*{Related literature}

The model connects the long-memory and asymmetric-volatility literatures. Hyperbolic
decay in absolute and squared returns motivates fractional specifications such as
FIGARCH \citep{ding1993,andersen1998,baillie1996}; HYGARCH, component GARCH, and
Markov-switching GARCH provide alternative paths to persistent volatility
\citep{davidson2004,engle1999,haas2004}. The symmetric model of \citet{lee2026} differs
by representing the persistence through a finite-dimensional Markov state consisting of
a variance level and a kernel-age coordinate, rather than an infinite-order fractional
recursion or latent regime. We retain that representation here, that makes the drift and
regularity analysis of Section~\ref{sec:theory} tractable.

Standard asymmetric models such as GJR-GARCH and EGARCH make the variance response
sign-dependent \citep{glosten1993,nelson1991}; FIAPARCH and FIEGARCH combine asymmetry
with fractional persistence \citep{tse1998,bollerslev1996}. In these specifications,
sign dependence acts primarily on response amplitude while the fractional decay order is
common across signs. ALM-GARCH instead lets shock sign reset the kernel-age coordinate
$\gamma_s$. Therefore, positive and negative shocks can have different 
persistence profiles in the finite-horizon while sharing the same asymptotic exponent $p$. The resulting
separation of amplitude from kernel-age persistence is therefore ALM-GARCH's specific contribution.

Because forecasts are evaluated against realized variance, Section~\ref{sec:emp_oos}
also includes HAR \citep{corsi2009}, a purpose-built realized-volatility benchmark, along
with return-based GARCH specifications.

Because the memory exponent $p$ is weakly identified separately from the kernel offset and it is
fixed by normalization (Section~\ref{sec:pident}). The sufficient drift certificate is
not obtained for one fitted configuration, and the forecast gains over the symmetric
long-memory model are concentrated in the DAX. Our focus therefore is on the
separation of volatility asymmetry into two testable channels within a finite-dimensional
Markov model.

The remainder of the paper is organized as follows. Section~\ref{sec:model} derives the
model. Section~\ref{sec:theory} establishes stationarity and positive Harris recurrence
under the active-branch and joint-stability conditions and gives the conditional
boundary-shift results, detailed proofs are deferred to the supplementary appendix.
Section~\ref{sec:identification} treats identification, the two nested tests, and the
information geometry of the two channels. Section~\ref{sec:empirics} presents the
empirical analysis. Section~\ref{sec:conclusion} sets the conclusion.


\section{The model}
\label{sec:model}

\subsection{The symmetric base model}
\label{sec:base}

We build on the long-memory GARCH (LM-GARCH) model of \citet{lee2026}, which we take
as the symmetric base. In that model the conditional variance is predictable,
\begin{equation}
\sigma_n^2 \;=\; \mu + X_{n-1},
\qquad \mu>0,
\label{eq:cond_var}
\end{equation}
where we impose a zero conditional mean for the observed log return $r_n$, so that
\begin{equation}
r_n\equiv\eta_n=\sigma_n\varepsilon_n,
\qquad U_n:=\varepsilon_n^2,
\qquad z_n:=\eta_n^2=\sigma_n^2 U_n,
\label{eq:standardized_innov}
\end{equation}
Let $\F_n$ denote the natural filtration generated by the returns and states up to time
$n$. The standardized innovations $\varepsilon_n$ are specified in
Assumption~\ref{ass:innov}. The pair $(X_n,c_n)$ evolves as a two-dimensional Markov
chain on $[0,\infty)\times[\gamma,\infty)$. The state $X_n$ carries the level of
accumulated variance and $c_n$ acts as a memory carrier, an effective kernel age that
compresses the past into a single additional coordinate. With $r(c)=(1+\tau/c)^{-1}$ and $\rho(c):=r(c)^p$, the update is
\begin{equation}
X_n = X_{n-1}\,r(c_{n-1})^p + \xi\, z_n,
\qquad
\frac{X_n}{c_n} = \frac{X_{n-1}}{c_{n-1}}\,r(c_{n-1})^{p+1} + \frac{\xi}{\gamma}\,z_n,
\label{eq:base_update}
\end{equation}
with parameters $\mu>0$, $\xi>0$, $\gamma>0$, $p>1$ and time step $\tau>0$. As shown
by \citet{lee2026}, the updates \eqref{eq:base_update} arise from matching the level
and slope at the origin of a power-law kernel $h_n(t)=\mu + a_n/(t+c_n)^p$ after each
shock injects a component $\alpha\,\eta_n^2/(t+\gamma)^p$, with $\xi := \alpha/\gamma^p$.
Here $X_n:=a_n/c_n^p$, so $h_n(0)=\mu+X_n$ and $h_n'(0)=-pX_n/c_n$; thus $X_n$ and $X_n/c_n$ represent, up to the fixed factor $-p$, the level and slope of the power-law component at the origin.
The chain reproduces hyperbolic-type memory decay while remaining finite-dimensional,
in contrast to the infinite-order representations of FIGARCH \citep{baillie1996} or
power-law Hawkes models.

\subsection{Sign-dependent kernel injection}
\label{sec:alm_def}

Empirically, the volatility response to returns is asymmetric: negative returns raise
future volatility by more than positive returns of the same magnitude
\citep{black1976,nelson1991,glosten1993}. The base model is symmetric in $\eta_n$ and
does not capture this. We introduce asymmetry at its structural source: by allowing the injected
kernel itself to depend on the sign of the shock.

Since $\sigma_n>0$, $\eta_n$ and $\varepsilon_n$ have the same sign. On a shock of
sign $s\in\{+,-\}$, we inject a
power-law component with its own amplitude and offset,
\begin{equation}
\frac{\alpha_s\,\eta_n^2}{(t+\gamma_s)^p},
\qquad
(\alpha_s,\gamma_s)=
\begin{cases}
(\alpha_+,\gamma_+), & \eta_n\ge 0,\\[2pt]
(\alpha_-,\gamma_-), & \eta_n< 0.
\end{cases}
\label{eq:sign_kernel}
\end{equation}
Running the level-and-slope matching of \citet{lee2026} with the sign-dependent
component \eqref{eq:sign_kernel} yields the asymmetric long-memory GARCH (ALM-GARCH)
recursion
\begin{equation}
X_n = X_{n-1}\,r(c_{n-1})^p + \xi_s\, z_n,
\qquad
\frac{X_n}{c_n} = \frac{X_{n-1}}{c_{n-1}}\,r(c_{n-1})^{p+1} + \frac{\xi_s}{\gamma_s}\,z_n,
\qquad \xi_s := \frac{\alpha_s}{\gamma_s^{\,p}},
\label{eq:alm_update}
\end{equation}
with $\xi_s=\xi_+$ when $\eta_n\ge0$ and $\xi_s=\xi_-$ when $\eta_n<0$. The conditional
variance remains predictable through \eqref{eq:cond_var}, so $\sigma_n^2$ is
$\F_{n-1}$ measurable and the timing of the base model is preserved.
Between arrival times the existing kernel ages deterministically. At each arrival,
shock sign selects the injected amplitude $\xi_s$ and reset offset $\gamma_s$.
Supplementary Figure~\ref{supp:fig:kernel} illustrates the mechanics of this update.

\subsection{Two channels of asymmetry}
\label{sec:channels}

The specification \eqref{eq:alm_update} contains two distinct departures from
symmetry, corresponding to the two free properties of the injected kernel
\eqref{eq:sign_kernel}: its amplitude and its offset.

The \emph{level channel} is the difference in injection amplitude. Writing
$\xi:=\xi_+$ and
\begin{equation}
\delta := \frac{\xi_-}{\xi_+}-1
= \frac{\alpha_-}{\alpha_+}\Big(\frac{\gamma_+}{\gamma_-}\Big)^{p}-1,
\label{eq:delta}
\end{equation}
the negative shock injection is $\xi_-=\xi(1+\delta)$. A positive $\delta$ therefore
captures an amplitude-based leverage-type asymmetry: negative shocks inject more variance
mass than positive shocks of the same squared magnitude. And this is the channel targeted, in
different functional form, by GJR-GARCH and EGARCH.

We use $s\in\{+,-\}$ for a generic sign branch, so subscripts such as $\xi_s$,
$\gamma_s$, $\rho_s$, and $\bar\kappa_s$ apply to either branch. Explicit $+$ and $-$
subscripts are used when we discuss the branches separately, including in the tables.
The parameterizations $(\xi_+,\xi_-)$ and $(\xi,\delta)$ are equivalent, with
$\xi:=\xi_+$ and $\delta$ defined in \eqref{eq:delta}. Estimation and reporting use
$(\xi_+,\xi_-)$ directly. We use $\delta$ for the leverage restriction $\delta=0$ and
for expressions, such as the envelope $S_\delta$ in Section~\ref{sec:theory}, that depend
on the overall asymmetry. For $k=0,1,\ldots$, define the sign-specific response profile
\[
g_s(k):=\xi_s\left(\frac{\gamma_s}{\gamma_s+k\tau}\right)^p.
\]
Persistence is summarized by the better-identified one-step persistence ratio
\[
\rho_s:=\frac{g_s(1)}{g_s(0)}
=\rho(\gamma_s)
=\left(\frac{\gamma_s}{\gamma_s+\tau}\right)^p.
\]
We use ``persistence summary'' and ``one-step persistence ratio'' interchangeably for
$\rho_s$.
The raw offsets $\gamma_s$ and the exponent $p$ are retained where they enter the model's
mechanics.

The \emph{memory channel} is the difference in offset, $\gamma_-\neq\gamma_+$. The
offset sets the age of the injected kernel; a shock that resets the kernel to a
different age changes its finite-horizon persistence profile independently of the
injection amplitude. Throughout, ``memory channel'' is shorthand for this difference in
finite-horizon persistence profiles. The common exponent $p$ fixes the asymptotic
power-law order for both signs, so different offsets do not imply different asymptotic
memory orders. This mechanism
differs from the amplitude-based asymmetry used in existing long-memory volatility
models discussed earlier in the Introduction section: sign dependence acts on the kernel-age coordinate
$\gamma_s$ and hence on the finite-horizon persistence profile, rather than only on the
amplitude of the variance response.

Amplitude and offset are separate properties of the kernel. Accordingly, $\delta$
isolates level asymmetry while holding the offset fixed, whereas $\gamma_-/\gamma_+$
isolates memory asymmetry. Section~\ref{sec:identification} therefore uses this separation to form
two nested tests. The distinction is structural and not statistical: near the symmetric
null, these two channels carry highly overlapping information, as discussed in
Section~\ref{sec:identification}.

\subsection{State space and nesting}
\label{sec:nesting}

Define $\gamma_{\min}:=\gamma_+\wedge\gamma_-$ and $\gamma_{\max}:=\gamma_+\vee\gamma_-$. Since the injected offset is either $\gamma_+$ or $\gamma_-$, the carrier $c_n$ is bounded below by $\gamma_{\min}$, and the chain $(X_n,c_n)$ lives on
\begin{equation}
\mathcal S = [0,\infty)\times[\gamma_{\min},\infty).
\label{eq:state_space}
\end{equation}
At $(\delta,\gamma_-)=(0,\gamma_+)$, equivalently $\xi_-=\xi_+$ and
$\gamma_-=\gamma_+$, the recursion \eqref{eq:alm_update} reduces exactly to the base
update \eqref{eq:base_update}. ALM-GARCH therefore nests the symmetric model of
\citet{lee2026}, and the restriction is testable. This exact nesting underlies both the
likelihood-ratio tests in Section~\ref{sec:identification} and the stability theory in
Section~\ref{sec:theory}, where every object reduces to its \citet{lee2026} counterpart
at the symmetric null.


\section{Stationarity, recurrence, and a diagnostic stability frontier}
\label{sec:theory}

For interior parameter values satisfying Assumption~\ref{ass:active}, we establish
stationarity and positive Harris recurrence of the ALM-GARCH chain $(X_n,c_n)$ under an
explicit condition, and then examine how the two asymmetry channels move a related
diagnostic frontier. This argument extends \citet{lee2026} to the
two-regime chain generated by sign-dependent injection. At
$(\xi_-,\gamma_-)=(\xi_+,\gamma_+)$, each object reduces to its symmetric counterpart.
We defer detailed proofs to the supplement appendix.

Throughout, write $r(c)=(1+\tau/c)^{-1}$ and $\rho(c):=r(c)^p$. Also,
$U=\varepsilon_n^2$ denotes the squared standardized innovation from
\eqref{eq:standardized_innov}; under the Gaussian reference law used for the
quasi-likelihood, $U\sim\chi^2(1)$. Since $\sigma_n>0$, define
\[q_-:=\Pr(\varepsilon_n<0)=\Pr(\eta_n<0),\qquad q_+:=1-q_-.\]
For $s\in\{+,-\}$, write $q_s$ for the corresponding branch probability. We maintain the following assumption, as in \citet{lee2026}.

\begin{assumption}
\label{ass:innov}
The standardized innovations $\{\varepsilon_n\}$ are i.i.d.\ with a lower
semicontinuous density that is strictly positive on $\mathbb R$ and symmetric about
zero. Hence $q_-=q_+=\tfrac12$, the sign of $\varepsilon_n$ is independent of
$U=\varepsilon_n^2$, and $U$ satisfies $\Pr(U>0)=1$, $\E[U]=1$, and
$\E[\log(1+U)]<\infty$.
\end{assumption}

\begin{assumption}
\label{ass:active}
Both injection amplitudes are strictly positive, $\xi_+>0$ and $\xi_->0$, so that both
sign branches are active and the injection ratios $\bar\kappa_s(c)=\gamma_s\,r(c)^{p+1}/
(\xi_s\,c)$ are finite. The recurrence and regularity results below are stated for
these interior parameter values. Estimation uses a log parameterization for both
injection amplitudes, so every reported fitted value is strictly positive; the very small
S\&P~500 and FTSE~100 positive-branch estimates are therefore near-boundary values, not
exact zeros. At the exact boundary $\xi_s=0$, the ratio-based envelope and the anchor
Jacobian used in the regularity proof degenerate and a separate boundary-state argument
would be required. We do not claim that extension here; empirically, proximity to that
boundary is handled through weak-identification language and floor-sensitivity checks.
\end{assumption}

\subsection{The asymmetric drift envelope}
\label{sec:drift}

The carrier dynamics are governed by the conditional one-step change in $c_n$. Because
the sign of the shock is, under Assumption~\ref{ass:innov}, independent of its
magnitude, the drift of the carrier is a two-component mixture, one component per sign.

\begin{lemma}[Asymmetric carrier-drift envelope]
\label{lem:envelope}
Under Assumptions~\ref{ass:innov} and~\ref{ass:active}, define, for
$c\ge\gamma_{\min}$ and $s\in\{+,-\}$,
\[
\bar\kappa_s(c)=\frac{\gamma_s r(c)^{p+1}}{\xi_s c},
\qquad
m(\kappa)=\E\!\left[\frac{\kappa}{\kappa+U}\right],
\]
and
\begin{equation}
S_\delta(c)
=
\sum_{s\in\{+,-\}}q_s
\left[-(c-\gamma_s)+(c+\tau-\gamma_s)
 m\!\left(\bar\kappa_s(c)\right)\right].
\label{eq:envelope}
\end{equation}
Writing
\[
D_c(x,c)=\E[c_n-c_{n-1}\mid X_{n-1}=x,c_{n-1}=c],
\]
the following statements hold:
\begin{enumerate}
\item[(i)] If $c+\tau-\gamma_s\ge0$ for both sign branches, then
\begin{equation}
D_c(x,c)\le S_\delta(c).
\label{eq:driftbound}
\end{equation}
\item[(ii)] For every $(x,c)\in\mathcal S$,
\[
D_c(x,c)\le \bar S,
\qquad
\bar S=\max\{\tau,\gamma_{\max}-\gamma_{\min}\}<\infty.
\]
\item[(iii)] For every compact
$K\subset[\gamma_{\min},\infty)$,
\[
\sup_{c\in K}|D_c(x,c)-S_\delta(c)|\longrightarrow0
\qquad\text{as }x\to\infty.
\]
\end{enumerate}
Moreover, $S_\delta(c)\le\bar S$ for all $c\ge\gamma_{\min}$ and
$S_\delta(c)\to-\infty$ as $c\to\infty$. 
At the symmetric case,
$S_\delta$ coincides with the carrier-drift envelope of
\citet{lee2026}.
\end{lemma}

The first bound in Lemma~\ref{lem:envelope} is a pointwise envelope on the
region $c\ge\gamma_{\max}-\tau$. On the complementary bounded strip, some
branch coefficients $c+\tau-\gamma_s$ may be negative, so $S_\delta(c)$ need
not be a finite-$x$ upper bound. What is needed for the Foster--Lyapunov proof
is instead the uniform convergence of the actual carrier drift to
$S_\delta$ on bounded $c$-sets, together with the global finite bound
$\bar S$. For large $c$, all branch coefficients are nonnegative and the
pointwise envelope applies directly. The supplement appendix illustrates the envelope and its bounded exceptional strip at the fitted Nikkei~225 configuration.

The envelope $S_\delta$ is the sign-weighted mixture of the two branch-specific
envelopes. For the Foster--Lyapunov argument, the mixture need only have a finite
upper bound and tend to $-\infty$ as $c\to\infty$. The sharper symmetric-model bound
$S_\delta\le\tau$ remains valid only when
$\gamma_{\max}-\gamma_{\min}\le\tau$; otherwise the finite bound $\bar S$ is used.

\subsection{Stationarity and positive Harris recurrence}
\label{sec:stationarity}

Let
\begin{equation}
\Lambda_\delta(c)
= \sum_{s\in\{+,-\}}q_s\,
\E\big[\log\!\big(\rho(c)+\xi_s\,U\big)\big]
\label{eq:lambda}
\end{equation}
denote the asymmetric far-field log-drift of $X_n$, the mixture counterpart of the
log-drift in \citet{lee2026}.

\begin{condition}[Joint stability]
\label{cond:stability}
There exists $\lambda>0$ such that
\begin{equation}
\sup_{c\ge\gamma_{\min}}\big\{\Lambda_\delta(c)+\lambda\,S_\delta(c)\big\}<0.
\label{eq:joint}
\end{equation}
\end{condition}

At the symmetric null, $\Lambda_\delta$ and $S_\delta$ reduce to their base-model forms
and \eqref{eq:joint} becomes the joint stability assumption of \citet{lee2026}.

\begin{proposition}[Regularity]
\label{prop:regularity}
Under Assumptions~\ref{ass:innov} and~\ref{ass:active}, there exists a nontrivial
measure $\phi$ such that the ALM-GARCH chain is $\phi$ irreducible, and every compact
subset of $\mathcal S$ is petite.
\end{proposition}

\begin{theorem}[Positive Harris recurrence]
\label{thm:stationarity}
Under Assumptions~\ref{ass:innov} and~\ref{ass:active} and
Condition~\ref{cond:stability}, the chain
$(X_n,c_n)$ on $\mathcal S=[0,\infty)\times[\gamma_{\min},\infty)$ is
positive Harris recurrent and admits a unique invariant
probability measure $\pi$. In particular, the strong law of large numbers holds for
$\pi$-integrable functionals of the chain.
\end{theorem}

The result establishes positive Harris recurrence, not geometric rates or aperiodicity.
Its proof combines the regularity result above with a Foster--Lyapunov drift for
$V(x,c)=\log(1+x)+\lambda c$; Lemma~\ref{lem:envelope} and
Condition~\ref{cond:stability} give negative drift outside a compact petite set, allowing
application of \citet[Theorem~11.3.4]{meyn2009markov}. The supplement appendix gives the
full control-map, minorization, and drift calculations.

We evaluate Condition~\ref{cond:stability} at the fitted parameters using deterministic
quadrature and an adaptively refined carrier grid. The diagnostic
$M^\ast=\inf_{\lambda>0}\sup_c\{\Lambda_\delta(c)+\lambda S_\delta(c)\}$ is negative for
five of the six fits (Table~\ref{tab:cond1}). For active-branch configurations, a negative
value verifies the sufficient condition. KOSPI is inside with little margin
($M^\ast=-0.000088$), whereas the Nikkei~225 has $M^\ast=0.003054$ and is not covered.
This is not evidence of instability; it demonstrates only that the sufficient log-linear
certificate lacks slack at that estimate. The S\&P~500 and FTSE~100 values are reported
as boundary-limit diagnostics because their positive branches are nearly inactive.
Supplementary Figures~\ref{supp:fig:cdrift} and~\ref{supp:fig:joint} report the diagnostic
curves for the fitted Nikkei~225 and the joint Bitcoin/Nikkei~225 comparison.

\begin{table}[hbt!]
\centering
\caption{Joint-stability certificate at the fitted $p=1.2$ configurations.}
\label{tab:cond1}
\begin{tabular}{lrrrl}
\toprule
Asset & $\lambda^\ast$ & $M^\ast$ & $c_{\arg\sup}$ & Certificate \\
\midrule
S\&P 500   & 0.0450 & $-0.011284$ & 11.63 & Boundary limit \\
FTSE 100   & 0.0469 & $-0.013400$ & 11.28 & Boundary limit \\
DAX        & 0.0341 & $-0.010012$ & 14.78 & Yes \\
Nikkei 225 & 0.0452 & $0.003054$  & 11.35 & Not covered \\
KOSPI      & 0.0312 & $-0.000088$ & 15.56 & Yes (borderline) \\
Bitcoin    & 0.0924 & $-0.016615$ &  5.72 & Yes \\
\bottomrule
\end{tabular}
\tabnote{$M^\ast=\inf_{\lambda>0}\sup_{c\ge\gamma_{\min}}
\{\Lambda_\delta(c)+\lambda S_\delta(c)\}$; $\lambda^\ast$ is the minimizing
weight and $c_{\arg\sup}$ is the carrier value attaining the reported supremum.
The calculations use deterministic Gauss--Hermite quadrature and an adaptively refined
carrier grid. A negative $M^\ast$ verifies Condition~\ref{cond:stability} only for
interior active-branch configurations satisfying Assumption~\ref{ass:active}. Values
for the S\&P~500 and FTSE~100 are boundary-limit diagnostics because the fitted
positive branch is nearly inactive. ``Not covered'' means that this sufficient
certificate was not obtained and does not imply instability; ``borderline'' denotes a
negative value with limited numerical margin.}
\end{table}

\subsection{A diagnostic stability frontier}
\label{sec:boundary}

We now describe how the two asymmetry channels move the diagnostic stability
frontier. Let $c_\delta^*$ denote the first zero of the corresponding
memory-scale drift envelope $S_\delta$, suppressing its dependence on $\xi$
and $\gamma_\pm$, and assume that it is a simple downward crossing:
\[
S_\delta(c_\delta^*)=0,
\qquad
S_\delta'(c_\delta^*)<0.
\]
Define the implied decay level
\[
\rho_\delta^* = \rho(c_\delta^*).
\]
The associated diagnostic is
\begin{equation}
	F(\xi,\delta;\gamma_+,\gamma_-)
	=
	q_+\,
	\E\!\left[
	\log\!\left(
	\rho_\delta^*+\xi U
	\right)
	\right]
	+
	q_-\,
	\E\!\left[
	\log\!\left(
	\rho_\delta^*+\xi(1+\delta)U
	\right)
	\right].
	\label{eq:diag}
\end{equation}
Its sign provides a diagnostic for the drift condition. Since
$S_\delta(c_\delta^*)=0$, negativity of $F$ at the crossing is necessary for
Condition~\ref{cond:stability} to hold. This diagnostic is necessary for the sufficient
certificate but not a necessary condition for stationarity itself. The corresponding diagnostic frontier
$\xi^*(\gamma_+,\gamma_-,\delta)$ is defined by
\[
F\bigl(
\xi^*(\gamma_+,\gamma_-,\delta),
\delta;
\gamma_+,\gamma_-
\bigr)
=
0,
\]
as in the $(\gamma,\xi)$ diagnostic of \citet{lee2026}.

For the symmetric model, define
\[
F_0(\xi)
:=
F(\xi,0;\gamma_+,\gamma_+),
\]
and let $\xi_0^*$ denote the corresponding frontier. The level channel affects
$F$ both directly, through the negative-shock amplitude, and indirectly,
through the induced movement of $c_\delta^*$. A global sign for the net effect
therefore requires the following dominance condition.

\begin{condition}[Derivative dominance]
\label{cond:dominance}
In the equal-reset case $\gamma_+=\gamma_-=\gamma$, let
$\rho'(c)=\rho(c)p\tau/[c(c+\tau)]$,
$\kappa_-:=\bar\kappa_-(c_\delta^*)$,
$m'(\kappa)=\E[U/(\kappa+U)^2]$, and
\[
W(c_\delta^*)=q_+\E\!\left[\frac{1}{\rho(c_\delta^*)+\xi U}\right]
+q_-\E\!\left[\frac{1}{\rho(c_\delta^*)+\xi(1+\delta)U}\right].
\]
At the relevant crossing require
\[
\E\!\left[\frac{U}{\rho(c_\delta^*)+\xi(1+\delta)U}\right]
\ge
\frac{\rho'(c_\delta^*)W(c_\delta^*)(c_\delta^*+\tau-\gamma)
\kappa_-m'(\kappa_-)}
{\xi(1+\delta)|S_\delta'(c_\delta^*)|}.
\]
\end{condition}

\begin{proposition}[Diagnostic-frontier decomposition: conditional and local form]
\label{prop:boundary}
Assume that the relevant diagnostic frontier is unique and that
$\partial F(\xi,\delta;\gamma_+,\gamma_-)/\partial\xi>0$ at the frontier.
\begin{enumerate}
\item[(i)] \emph{Level channel.} At the symmetric null,
\begin{equation}
\left.\frac{\partial}{\partial\delta}F(\xi,\delta;\gamma_+,\gamma_+)
\right|_{\delta=0}=q_-\xi\,\frac{\D}{\D\xi}F_0(\xi).
\label{eq:localnull}
\end{equation}
Hence $F_0'(\xi)>0$ makes a small positive level asymmetry locally destabilizing. More
generally, wherever Condition~\ref{cond:dominance} holds,
$\partial F/\partial\delta\ge0$ and $\xi^*(\gamma_+,\gamma_+,\delta)$ is nonincreasing in
$\delta$.
\item[(ii)] \emph{Memory channel.} With $\delta=0$, if
$c_0^*+\tau-\gamma_-\ge0$, lowering $\gamma_-$ below $\gamma_+$ decreases
$F(\xi,0;\gamma_+,\gamma_-)$ at a given $\xi$, so
$\xi^*(\gamma_+,\gamma_-,0)\ge\xi_0^*$.
\item[(iii)] \emph{Bracket.} In the equal-reset case, if
$F(\xi_0^*,\delta;\gamma_+,\gamma_+)\ge0$ and
$F(\xi_0^*/(1+\delta),\delta;\gamma_+,\gamma_+)\le0$, then
\begin{equation}
\frac{\xi_0^*}{1+\delta}\le\xi^*(\gamma_+,\gamma_+,\delta)\le\xi_0^*.
\label{eq:bracket}
\end{equation}
\end{enumerate}
\end{proposition}

The local result \eqref{eq:localnull} does not require
Condition~\ref{cond:dominance}; as the latter controls the indirect movement of
$c_\delta^*$ needed for a global level-channel sign. Numerically, the condition holds
with roughly a threefold margin for the four fitted configurations with an active,
separately estimable positive branch (DAX, Nikkei, KOSPI, and Bitcoin). It is not
applicable to the near-boundary S\&P~500 and FTSE~100 fits and is not asserted over the
full parameter space. Under Proposition~\ref{prop:boundary}, level asymmetry moves the
diagnostic frontier inward while a harder negative-shock reset moves it outward; their
net effect is therefore empirical.


\section{Identification and estimation}
\label{sec:identification}

We first describe the quasi-likelihood estimation and then discuss two features that
affect inference. The exponent $p$ is weakly identified separately from the kernel offset,
so it is fixed rather than estimated. In addition, the level and memory channels are
structurally distinct but carry nearly the same local information near the symmetric
null. Both issues influence the interpretation of the empirical results.

\subsection{Estimation}
\label{sec:estimation}

We estimate parameters by Gaussian quasi-maximum likelihood. Because
$\sigma_n^2=\mu+X_{n-1}$ is predictable, the period-$n$ return is evaluated before the
state update. We estimate $(\xi_+,\xi_-)$ directly on the log scale, which is equivalent
to the $(\xi,\delta)$ parameterization but remains convenient near an inactive branch.
The main sections of the paper report point estimates rather than Wald inference. Sandwich-robust
standard errors based on the covariance $n^{-1}H^{-1}JH^{-1}$ are reported in the supplement appendix where
identification is regular; boundary configurations do not support meaningful
covariance-based inference. The channel conclusions therefore rely on the bootstrap
likelihood-ratio tests.

\subsection{Weak identification of the memory exponent}
\label{sec:pident}

In the symmetric base model, the offset $\gamma$ and exponent $p$ jointly determine the persistence of the volatility
kernel. In the recursion they enter mainly through the one-step decay factor
\begin{equation}
\rho=\rho(\gamma)=\Big(\frac{\gamma}{\gamma+\tau}\Big)^{p}.
\label{eq:rhostar}
\end{equation}
Consequently, many $(p,\gamma)$ pairs produce nearly the same likelihood when they imply
similar values of $\rho$. The profile likelihood is consistent with this weak
separation. Across
$p\in[1.05,3.0]$, the maximized log-likelihood changes by at most $8.4$ points and by
less than five points for four of the six assets. Over the same range, $\rho$ changes
by only $0.05$--$0.08$ on the equity indices even when $p$ more than doubles. When $p$
is estimated freely, it reaches the imposed boundary for several assets, while the
persistence summary remains stable.

We therefore fix $p=1.2$, the value used in the symmetric base model of
\citet{lee2026}, so that the two specifications remain directly comparable. Relative to
the profile optimum, the loss in log-likelihood is at most $6.6$ points and less than
four points in four markets. Section~\ref{sec:emp_robust} repeats the analysis at
$p=2.0$ and obtains the same substantive conclusions. The parameter interpreted below is
the better-identified persistence composite $\rho$, rather than $p$ alone.

\subsection{Geometry of the two channels}
\label{sec:channel_geom}

The two asymmetry channels correspond to the two free properties of the injected kernel.
The level channel is the difference in injection amplitude, summarized by
$\delta=\xi_-/\xi_+-1$; the memory channel is the difference in offset,
$\gamma_-\neq\gamma_+$. Geometrically these are independent: $\delta$ scales the
magnitude of the negative-shock response while leaving its kernel age fixed, whereas
$\gamma_-/\gamma_+$ changes the age of the negative-shock kernel while leaving its
magnitude fixed. Thus the \emph{level-only} case is $\delta\neq0$ with
$\gamma_-=\gamma_+$, whereas the \emph{memory-only} case is $\delta=0$ with
$\gamma_-\neq\gamma_+$. The former produces amplitude-based leverage-type asymmetry
without sign-dependent finite-horizon persistence; the latter produces sign-dependent
finite-horizon persistence without amplitude asymmetry. The two are distinct features of
the volatility response, and the second, sign-dependent kernel aging, is the mechanism
without an amplitude-based analogue in the GARCH family discussed earlier in the Introduction section.
Because $p$ is common across branches, this distinction concerns finite-horizon
persistence and does not imply sign-specific asymptotic memory orders.

\subsection{Nested tests and information geometry}
\label{sec:tests}

We test each channel by a likelihood-ratio statistic against the corresponding restricted
model: $LR_{\mathrm{lev}}$ for $\xi_-=\xi_+$, $LR_{\mathrm{mem}}$ for $\gamma_-=\gamma_+$,
and a joint statistic $LR_{\mathrm{sym}}$ for both restrictions simultaneously. Under
a correctly specified Gaussian likelihood and regular interior identification,
conventional likelihood-ratio theory gives $\chi^2$ reference distributions with one,
one, and two degrees of freedom. Because estimation is used here as Gaussian
quasi-likelihood and these regularity conditions are not established for every fitted
configuration, we treat the $\chi^2$ values as reference calibrations rather than as the
basis for the main inference \citep{white1982}.

The channel specific tests are closely related near the symmetric null. Although the
underlying parameters describe different features of the kernel, the scores for
$\delta$ and the memory offset have a correlation of about $0.95$ in the Monte Carlo
design. Locally, a small change in amplitude and a small change in offset alter the early
part of the kernel in almost proportional ways. The two effects become distinguishable
only farther from the null. The joint statistic $LR_{\mathrm{sym}}$ is not the sum of
the channel specific
statistics. Each individual test has low local power, so non-rejection near the null
provides limited evidence that a channel is absent. We therefore report the joint test as
the main test of asymmetry and use the channel specific tests where the data contain
enough variation to separate the two effects. In Section~\ref{sec:empirics}, the
memory channel rejections occur for the assets with larger estimated offset asymmetry.

\subsection{Finite sample calibration of the tests}
\label{sec:test_reliability}

Restricted fits are obtained from multiple starting values for both the null and
unrestricted models; the retained optima are stable across starts. This matters most when
one branch is nearly inactive, where a poor restricted optimum can substantially inflate
$LR_{\mathrm{mem}}$.

The $\chi^2$ reference also requires qualification. We establish positive Harris
recurrence and an ergodic law of large numbers, but not the stronger conditions needed
for a full score limit theory. Near the symmetric null the information matrix is almost
singular, and for the S\&P~500 and FTSE~100 $\hat\xi_+\!\approx\!0$, placing the memory
offset near a boundary where the usual $\chi^2_1$ law need not apply
\citep{chernoff1954,self1987,andrews2001}. A representative interior null has empirical
sizes $0.060$, $0.053$, and $0.056$ for the level, memory, and joint tests, but the fitted
boundary-null monte-carlo runs in Section~\ref{sec:emp_mc} show $20$--$28\%$ rejection for
the conventional memory test. We therefore use finite sample bootstrap inference and
treat non-rejection in the two inactive-branch markets as weak identification rather
than evidence of equal offsets.

For each restriction, we fit the restricted null, simulate $B$ samples, and refit both
models using the same multi-start procedure. We use $B=999$, except $B=3{,}999$ for
DAX memory cell; the add-one $p$-value is
$(1+\#\{LR_b\ge LR_{\mathrm{obs}}\})/(B+1)$. All reported replications converged. We
also repeat the design with symmetrized standardized residuals in place of Gaussian
innovations. Table~\ref{tab:boot} reports both calibrations, which give the same
five-percent classifications. We provide the full simulation and convergence details in the supplement appendix.


\section{Empirical analysis}
\label{sec:empirics}

We estimate ALM-GARCH on five international equity indices and Bitcoin. The empirical
analysis asks whether the two asymmetry channels are separately identified and whether
the asymmetric extension adds fit or forecasting value relative to the symmetric model
and established volatility benchmarks.

\subsection{Data and estimation}
\label{sec:emp_data}

For the S\&P~500, FTSE~100, DAX, Nikkei~225, and KOSPI, we use daily open-to-close log
returns and five-minute realized variance from the Oxford--Man Institute Realized
Library \citep{heber2009}. The samples run from January~2000 to June~27, 2018, subject to
the first
available observation for each market. For Bitcoin, we use UTC-day close-to-close log
returns and one-minute realized variance constructed from Bitstamp one-minute prices
from January~2, 2012 to January~6, 2025. Missing minutes on a complete UTC grid are
filled by the previous close, so they contribute zero one-minute return. The exact sample counts and construction details are reported in Supplementary Table~\ref{supp:tab:data}. We treat Bitcoin as an additional application rather than a homogeneous cross sectional observation since it differs in
trading calendar, source, realized-variance frequency, and sample window.
The conditional mean is fixed at zero, so the observed log returns are the $\eta_n$
sequence in \eqref{eq:standardized_innov}. Returns enter the Gaussian quasi-likelihood;
realized variance is reserved for out-of-sample evaluation.

We estimate by quasi-maximum likelihood at fixed $p=1.2$. The injection amplitudes are
estimated directly as $(\xi_+,\xi_-)$, and persistence is reported through the
one-step persistence ratios $\rho_\pm=g_\pm(1)/g_\pm(0)
=(\gamma_\pm/(\gamma_\pm+\tau))^p$. The value $p=1.2$ follows the symmetric
model of \citet{lee2026}. Profile likelihoods are flat over nearby values, while the
persistence summaries are comparatively stable; Supplementary Figure~\ref{supp:fig:profile}
and Table~\ref{supp:tab:profile} report the full profile evidence. We treat $p$ as an identifying normalization rather than
as a precisely estimated shape parameter. Section~\ref{sec:emp_robust} reports the
corresponding channel results at $p=2.0$.

\begin{figure}[hbt!]
\centering
\includegraphics[width=\linewidth]{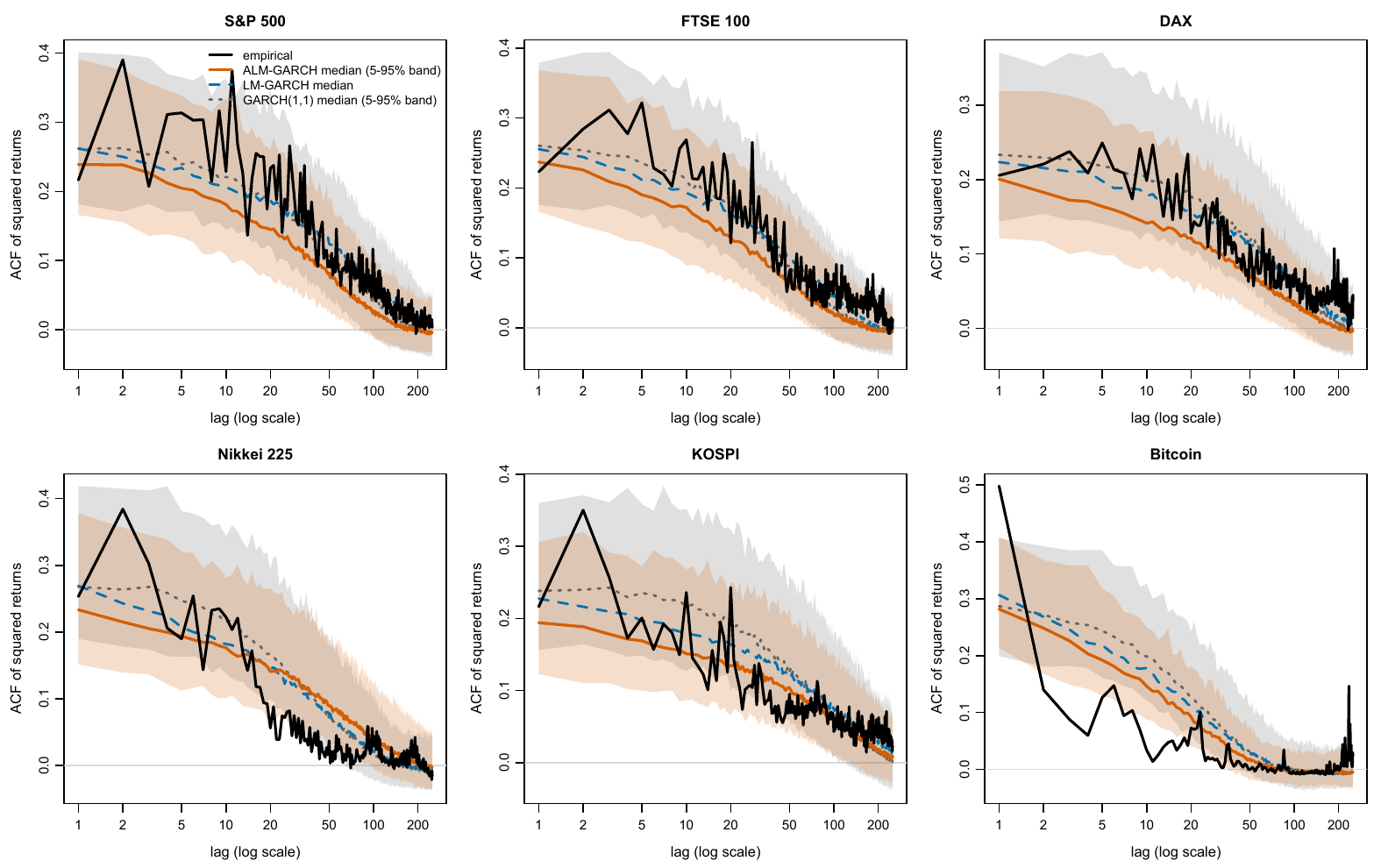}
\caption{\figACFCaption}
\label{fig:acf_persistence_fit}
\end{figure}

Figure~\ref{fig:acf_persistence_fit} compares squared-return autocorrelations with
simulation summaries from the fitted ALM-GARCH, LM-GARCH, and GARCH$(1,1)$ models.
The simulation profiles overlap over most lags, and ALM-GARCH and LM-GARCH are close
throughout. The figure documents consistency with the observed persistence profile,
but it does not sharply discriminate among highly persistent models at these sample
lengths.

\FloatBarrier

\subsection{Tests of the asymmetry channels}
\label{sec:emp_channels}

We test the level restriction $\xi_-=\xi_+$, the memory restriction
$\gamma_-=\gamma_+$, and the joint symmetry restriction. Table~\ref{tab:main} reports
the multi-start likelihood-ratio statistics. Because standard likelihood-ratio
references can be unreliable near inactive-branch boundaries, the principal
finite-sample conclusions use the fixed-replication bootstrap in
Table~\ref{tab:boot}.

\begin{table}[hbt!]
\centering
\caption{Likelihood-ratio tests of the asymmetry channels at fixed $p=1.2$.}
\label{tab:main}
\begin{tabular}{lrrrrrr}
\toprule
Asset & $LR_{\mathrm{lev}}$ & $p_{\mathrm{lev}}$ & $LR_{\mathrm{mem}}$ & $p_{\mathrm{mem}}$ & $LR_{\mathrm{sym}}$ & $p_{\mathrm{sym}}$ \\
\midrule
S\&P 500 & 110.01 & $<0.001$ & 0.15 & 0.699 & 187.09 & $<0.001$ \\
FTSE 100 & 101.64 & $<0.001$ & 0.94 & 0.331 & 210.87 & $<0.001$ \\
DAX & 87.72 & $<0.001$ & 7.23 & 0.007 & 157.66 & $<0.001$ \\
Nikkei 225 & 32.06 & $<0.001$ & 15.01 & $<0.001$ & 38.38 & $<0.001$ \\
KOSPI & 49.43 & $<0.001$ & 26.08 & $<0.001$ & 49.86 & $<0.001$ \\
Bitcoin & 16.62 & $<0.001$ & 24.82 & $<0.001$ & 25.83 & $<0.001$ \\
\bottomrule
\end{tabular}

\tabnote{$LR_{\mathrm{lev}}$ tests the level restriction $\xi_-=\xi_+$,
$LR_{\mathrm{mem}}$ tests the memory restriction $\gamma_-=\gamma_+$, and
$LR_{\mathrm{sym}}$ tests both restrictions jointly. The reported $p$-values use the
standard asymptotic likelihood-ratio reference distributions; values below 0.001 are
reported as $p<0.001$. All unrestricted and
restricted specifications use the same multi-start estimation procedure. Because
inactive-branch boundaries can invalidate the asymptotic reference, Table~\ref{tab:boot}
provides the primary finite-sample calibration.}
\end{table}

\begin{table}[hbt!]
\centering
\caption{Bootstrap calibration of the channel likelihood-ratio tests at fixed $p=1.2$.}
\label{tab:boot}
\begin{tabular}{lrrrrrrrrr}
\toprule
 & \multicolumn{3}{c}{$LR_{\mathrm{lev}}$} & \multicolumn{3}{c}{$LR_{\mathrm{mem}}$} & \multicolumn{3}{c}{$LR_{\mathrm{sym}}$} \\
\cmidrule(lr){2-4}\cmidrule(lr){5-7}\cmidrule(lr){8-10}
Asset & statistic & $p_{\mathrm{boot}}$ & $p_{\mathrm{boot}}^{\mathrm{res}}$ & statistic & $p_{\mathrm{boot}}$ & $p_{\mathrm{boot}}^{\mathrm{res}}$ & statistic & $p_{\mathrm{boot}}$ & $p_{\mathrm{boot}}^{\mathrm{res}}$ \\
\midrule
S\&P 500 & 110.01 & 0.001 & 0.001 & 0.15 & 0.853 & 0.890 & 187.09 & 0.001 & 0.001 \\
FTSE 100 & 101.64 & 0.001 & 0.001 & 0.95 & 0.573 & 0.646 & 210.87 & 0.001 & 0.001 \\
DAX & 87.72 & 0.001 & 0.001 & 7.23 & 0.066 & 0.141 & 157.65 & 0.001 & 0.001 \\
Nikkei 225 & 32.06 & 0.001 & 0.003 & 15.01 & 0.001 & 0.046 & 38.38 & 0.001 & 0.007 \\
KOSPI & 49.42 & 0.001 & 0.001 & 26.07 & 0.001 & 0.001 & 49.85 & 0.001 & 0.001 \\
Bitcoin & 16.62 & 0.001 & 0.035 & 24.82 & 0.001 & 0.011 & 25.83 & 0.001 & 0.027 \\
\bottomrule
\end{tabular}

\tabnote{The columns labelled ``statistic'' reproduce the observed likelihood-ratio
statistics. $p_{\mathrm{boot}}$ is the Gaussian parametric-bootstrap $p$-value and
$p_{\mathrm{boot}}^{\mathrm{res}}$ is the symmetrized-residual bootstrap value. Each
bootstrap statistic is recomputed after the restricted null is fitted using the same
multi-start procedure as for the observed data. The add-one calculation is
$(1+\text{exceedances})/(B+1)$. The design uses $B=999$ replications for every cell
except the DAX memory test, for which $B=3{,}999$.}
\end{table}

Complete symmetry is rejected in all six series, but this joint result does not imply
that both channel restrictions are rejected separately in every market. The level
restriction is strongly rejected throughout. The bootstrap memory test rejects for
Nikkei~225, KOSPI, and Bitcoin, but not for DAX at the five-percent level. For the
S\&P~500 and FTSE~100, the estimated positive-branch amplitude is nearly zero, so the
positive-branch offset is weakly identified. Floor-constrained refits in Supplementary Table~\ref{supp:tab:floor} do not produce a
memory-channel rejection in those two markets. Supplementary Figure~\ref{supp:fig:crossmarket}
summarizes the cross-market evidence for both asymmetry channels. We therefore describe
the evidence as pervasive level asymmetry and more selective
sign-dependent memory, rather than treating non-rejection as evidence that the memory
parameters are equal.

\begin{figure}[hbt!]
\centering
\includegraphics[width=\linewidth]{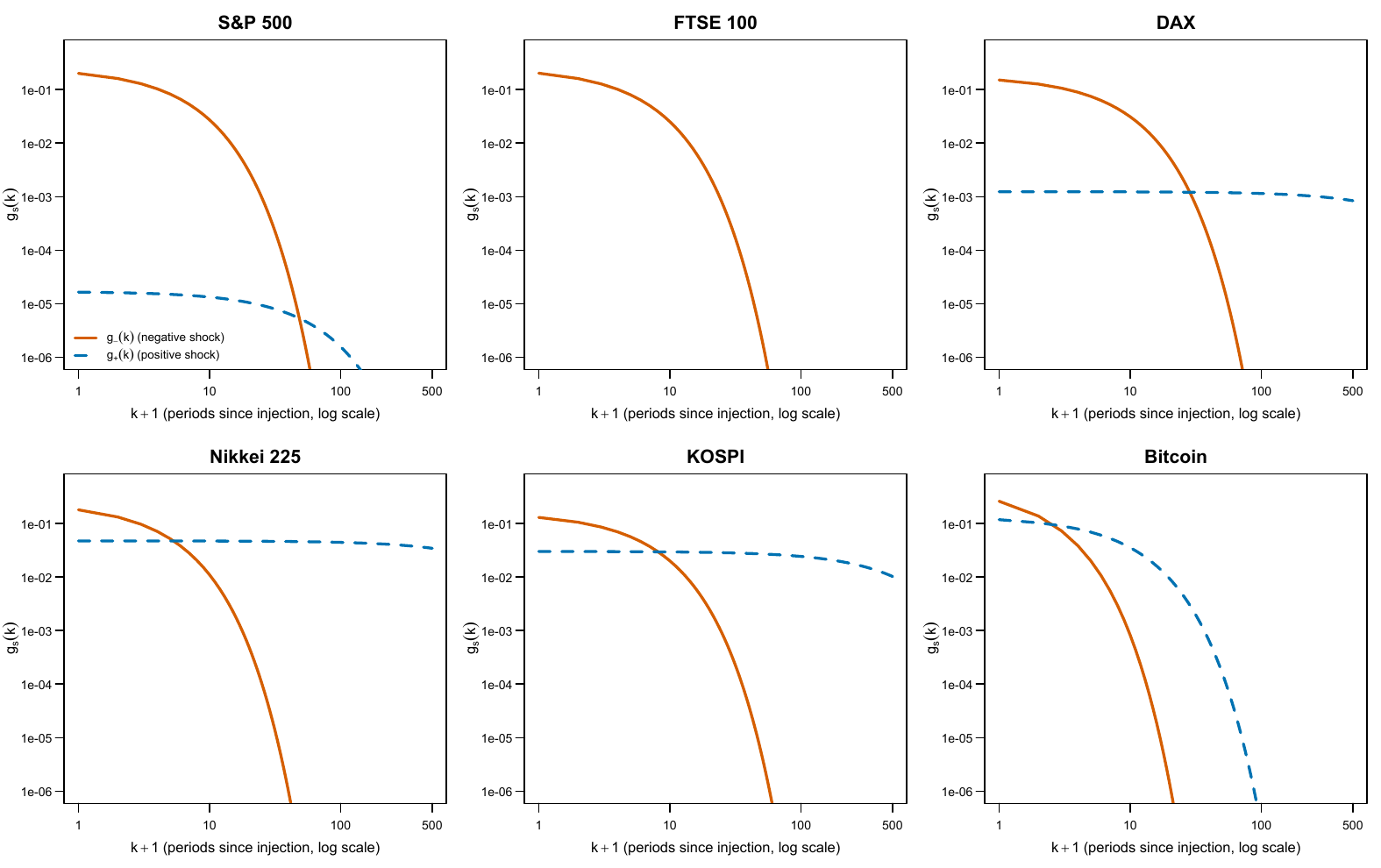}
\caption{\figResponseCaption}
\label{fig:sign_shock_responses}
\end{figure}

Figure~\ref{fig:sign_shock_responses} displays the fitted response
$g_s(k)=\xi_s\{\gamma_s/(\gamma_s+k\tau)\}^{p}$ for each sign. The vertical gap at
$k=0$ is the level channel, while differences in the subsequent finite-horizon profile
represent the offset-driven memory channel. Both branches share the same asymptotic
power exponent $p$. In the markets with a memory-channel rejection, positive-shock
responses are smaller initially but remain more persistent over the plotted horizon. The
nearly inactive positive branches for the
S\&P~500 and FTSE~100 demonstrate directly why their memory offsets are weakly identified.

\FloatBarrier

\subsection{In-sample fit against alternative volatility benchmarks}
\label{sec:insample}

Table~\ref{tab:insample} compares ALM-GARCH with the symmetric LM-GARCH model,
GARCH$(1,1)$, GJR-GARCH, and component GARCH under a common Gaussian
quasi-likelihood convention. ALM-GARCH improves on the symmetric long-memory model in
sample, but the short-memory asymmetric models remain competitive. The comparison is
therefore evidence that the asymmetric decomposition is empirically relevant, not that
ALM-GARCH universally dominates standard alternatives.

\begin{table}[hbt!]
\centering
\caption{In-sample fit against low-dimensional volatility benchmarks.}
\label{tab:insample}
\begin{tabular}{@{}lrrrrr@{}}
\toprule
& \multicolumn{5}{c}{Model} \\
\cmidrule(lr){2-6}
Asset & LM & GARCH$(1,1)$ & GJR & csGARCH & ALM \\
\midrule
\multicolumn{6}{@{}l}{\textit{Panel A: BIC}} \\
S\&P 500 & $-39175.6$ & $-39180.9$ & $\mathbf{-39358.9}$ & $-39192.1$ & $-39345.8$ \\
FTSE 100 & $-38887.4$ & $-38886.4$ & $\mathbf{-39082.6}$ & $-38896.7$ & $-39081.3$ \\
DAX & $-37870.2$ & $-37882.4$ & $\mathbf{-38017.1}$ & $-37874.1$ & $-38010.9$ \\
Nikkei 225 & $-37076.0$ & $-37091.1$ & $\mathbf{-37114.3}$ & $-37099.5$ & $-37097.6$ \\
KOSPI & $-37844.5$ & $-37853.4$ & $-37860.5$ & $-37873.3$ & $\mathbf{-37877.4}$ \\
Bitcoin & $-27263.7$ & $-27256.6$ & $-27248.6$ & $\mathbf{-27281.2}$ & $-27272.6$ \\
\addlinespace[3pt]
\multicolumn{6}{@{}l}{\textit{Panel B: Quasi-log-likelihood}} \\
S\&P 500 & $19600.5$ & $19603.1$ & $\mathbf{19696.3}$ & $19617.2$ & $19694.0$ \\
FTSE 100 & $19456.4$ & $19455.9$ & $19558.2$ & $19469.5$ & $\mathbf{19561.8}$ \\
DAX & $18947.8$ & $18953.9$ & $19025.5$ & $18958.2$ & $\mathbf{19026.6}$ \\
Nikkei 225 & $18550.6$ & $18558.1$ & $\mathbf{18574.0}$ & $18570.8$ & $18569.8$ \\
KOSPI & $18934.9$ & $18939.3$ & $18947.1$ & $18957.7$ & $\mathbf{18959.8}$ \\
Bitcoin & $13644.5$ & $13641.0$ & $13641.2$ & $\mathbf{13661.8}$ & $13657.4$ \\
\bottomrule
\end{tabular}
\tabnote{Panel~A reports the Bayesian information criterion (BIC), for which lower
values indicate better fit after penalizing parameter count. Panel~B reports the
maximized Gaussian quasi-log-likelihood, for which higher values indicate better fit.
Bold identifies the best value within each market and panel. LM denotes the symmetric
LM-GARCH model, GJR denotes GJR-GARCH, csGARCH denotes component GARCH, and ALM denotes
ALM-GARCH. The LM-GARCH and ALM-GARCH entries use fixed $p=1.2$.}
\end{table}

FIAPARCH is the more demanding asymmetric long-memory benchmark. The implementation
uses a $5{,}000$ lag ARCH$(\infty)$ approximation and multiple starting values to reduce
sensitivity to local optima.
Table~\ref{tab:fiaparch} shows that FIAPARCH has the higher quasi-log-likelihood and
lower BIC in all six series. ALM-GARCH is therefore not the best-fitting asymmetric
long-memory model in this comparison. The Model's distinct role is the separately testable
decomposition of response amplitude and finite-horizon persistence within a finite-dimensional
Markov representation.

\begin{table}[hbt!]
\centering
\caption{In-sample comparison with FIAPARCH$(1,d,1)$.}
\label{tab:fiaparch}
\begin{tabular}{@{}lrrrrrr@{}}
\toprule
 & \multicolumn{3}{c}{Quasi-log-likelihood} & \multicolumn{3}{c}{BIC} \\
\cmidrule(lr){2-4}\cmidrule(lr){5-7}
Asset & FIAPARCH & ALM & $\Delta$ & FIAPARCH & ALM & $\Delta$ \\
\midrule
S\&P 500 & $19750.2$ & $19694.0$ & $+56.2$ & $-39449.8$ & $-39345.8$ & $-104.0$ \\
FTSE 100 & $19578.8$ & $19561.8$ & $+17.0$ & $-39106.8$ & $-39081.3$ & $-25.5$ \\
DAX & $19047.5$ & $19026.6$ & $+20.9$ & $-38044.3$ & $-38010.9$ & $-33.4$ \\
Nikkei 225 & $18608.3$ & $18569.8$ & $+38.5$ & $-37166.2$ & $-37097.6$ & $-68.6$ \\
KOSPI & $18997.4$ & $18959.8$ & $+37.6$ & $-37944.2$ & $-37877.4$ & $-66.8$ \\
Bitcoin & $13682.0$ & $13657.4$ & $+24.6$ & $-27313.3$ & $-27272.6$ & $-40.7$ \\
\bottomrule
\end{tabular}
\tabnote{Both models are evaluated on the same observations under the Gaussian
quasi-likelihood convention. $\Delta$ is calculated as FIAPARCH minus ALM-GARCH within
each metric. Consequently, a positive quasi-log-likelihood difference and a negative
BIC difference both favor FIAPARCH. FIAPARCH has six estimated parameters; ALM-GARCH
has five when $p=1.2$ is fixed.}
\end{table}

\FloatBarrier

\subsection{Monte Carlo evidence}
\label{sec:emp_mc}

The representative Monte Carlo design and the fitted-null calibration answer different
questions. Under the representative interior symmetric null, empirical size is close to
five percent. At the fitted memory-restricted nulls for the S\&P~500, FTSE~100, and DAX,
the positive branch is boundary-prone and the conventional asymptotic memory test rejects
a true null about $20$--$25\%$ of the time. By contrast, the unrestricted fitted model
has four active positive branches (DAX, Nikkei~225, KOSPI, and Bitcoin); the separate
unrestricted fitted-design experiment identifies the S\&P~500 and FTSE~100 as the two
near-boundary configurations in which positive-branch persistence is poorly recovered.
These results motivate bootstrap calibration and the weak-identification convention.
Supplementary Tables~\ref{supp:tab:mcsize}, \ref{supp:tab:nullsize}, and~\ref{supp:tab:mccalib}
report, respectively, the representative-null size and power, fitted restricted-null size,
and unrestricted fitted-design calibration.

\FloatBarrier

\subsection{Estimation under the stability certificate}
\label{sec:emp_stab}

Given Condition~\ref{cond:stability} is sufficient rather than necessary. The unconstrained
Nikkei~225 estimate is just outside the verified region, while the KOSPI estimate is
inside with little margin. Table~\ref{tab:stabrefit} re-estimates the model subject to
$M^\ast(\theta)\le-10^{-3}$. The constraint costs $0.61$ log-likelihood points for
Nikkei~225 and $0.05$ for KOSPI, moves every channel likelihood-ratio statistic by less
than $1.3$ points, and changes no channel classification. The missing certificate for
the unconstrained Nikkei fit is therefore evidence that the sufficient certificate has
little slack at that estimate, not evidence that the fitted process is unstable.

\begin{table}[hbt!]
\centering
\caption{Stability-constrained ALM-GARCH refits at fixed $p=1.2$.}
\label{tab:stabrefit}
\begin{tabular}{lcrrrrrr}
\toprule
Asset & Binds & $\log L_{\mathrm{unres.}}$ & $\Delta\log L$ & $M^*$ before & $M^*$ after & $\hat\rho_+$ & $\hat\rho_-$ \\
\midrule
S\&P 500 & No & 19694.01 & 0.00 & $-1.1\times10^{-2}$ & $-1.1\times10^{-2}$ & 0.977 & 0.801 \\
FTSE 100 & No & 19561.78 & 0.00 & $-1.3\times10^{-2}$ & $-1.3\times10^{-2}$ & 0.986 & 0.793 \\
DAX & No & 19026.58 & 0.00 & $-1.0\times10^{-2}$ & $-1.0\times10^{-2}$ & 0.999 & 0.839 \\
Nikkei 225 & Yes & 18569.81 & $-0.61$ & $+3.1\times10^{-3}$ & $-1.0\times10^{-3}$ & 0.999 & 0.740 \\
KOSPI & Yes & 18959.77 & $-0.05$ & $-8.8\times10^{-5}$ & $-1.0\times10^{-3}$ & 0.999 & 0.816 \\
Bitcoin & No & 13657.45 & 0.00 & $-1.7\times10^{-2}$ & $-1.7\times10^{-2}$ & 0.875 & 0.528 \\
\bottomrule
\end{tabular}

\tabnote{The model is re-estimated subject to $M^\ast(\theta)\le -10^{-3}$.
``Binds'' indicates that the unrestricted estimate did not satisfy this margin and was
therefore refitted. ``Unrestricted $\log L$'' is the original maximized
quasi-log-likelihood and $\Delta\log L$ is constrained minus unrestricted, so a negative
value is the fit cost of the constraint. $M^\ast$ before and after are the sufficient
certificate diagnostics. The after values for the Nikkei~225 and KOSPI lie on the
imposed certificate boundary at the displayed precision. $\hat\rho_+$ and
$\hat\rho_-$ are the fitted one-period persistence summaries. Final solutions are
verified with the full certificate search using the parameter rounding of
Table~\ref{tab:cond1}.}
\end{table}

\FloatBarrier

\subsection{Out-of-sample forecast comparison}
\label{sec:emp_oos}

We hold out the final $30\%$ of each sample and form one-step-ahead forecasts by
expanding window estimation with re-estimation every $250$ days. Forecasts are scored
against the market-specific realized-variance target using QLIKE
\citep{patton2011}: five-minute realized variance for the equity indices and one-minute
realized variance for Bitcoin. All models use the same finite evaluation dates. For each
benchmark, the Diebold--Mariano statistic \citep{dieboldmariano1995} is based on
$d_t=L_t^{\mathrm{ALM}}-L_t^{\mathrm{benchmark}}$, so negative values favor ALM-GARCH,
with a Newey--West/Bartlett long-run variance and lag
$\lfloor n_{\mathrm{eval}}^{1/3}\rfloor$ \citep{neweywest1987}. Per-date forecasts and
date checks are retained in the replication materials. Table~\ref{tab:oos} reports the
common-date QLIKE comparison.

\begin{table}[hbt!]
\centering
\caption{Out-of-sample QLIKE losses at fixed $p=1.2$.}
\label{tab:oos}
\begin{tabular}{lrrrrrrr}
\toprule
Asset & ALM & LM-GARCH & HAR-RV & GARCH & GJR & EGARCH & FIGARCH \\
\midrule
S\&P 500 & 0.363 & 0.373 & 0.245 & 0.377 & 0.357 & 0.335 & 0.366 \\
FTSE 100 & 0.266 & 0.281 & 0.253 & 0.281 & 0.265 & 0.272 & 0.273 \\
DAX & 0.203 & 0.221 & 0.175 & 0.221 & 0.212 & 0.204 & 0.219 \\
Nikkei 225 & 0.367 & 0.374 & 0.293 & 0.369 & 0.368 & 0.333 & 0.362 \\
KOSPI & 0.207 & 0.208 & 0.141 & 0.213 & 0.228 & 0.218 & 0.173 \\
Bitcoin & 0.342 & 0.336 & 0.280 & 0.336 & 0.340 & 0.327 & 0.312 \\
\bottomrule
\end{tabular}

\tabnote{Lower QLIKE indicates greater forecast accuracy. The evaluation window is the
final $30\%$ of each sample, with expanding-window re-estimation every $250$ days. The
target is five-minute realized variance for the equity indices and one-minute realized
variance for Bitcoin. ALM denotes ALM-GARCH; LM-GARCH is the symmetric long-memory
model; HAR-RV is the heterogeneous autoregressive model of realized volatility; GJR is
GJR-GARCH; and EGARCH and FIGARCH retain their standard meanings. All seven models are
evaluated on identical finite-loss dates within each market.}
\end{table}

The return-based models are broadly comparable rather than uniformly ordered. ALM-GARCH
is significantly more accurate than GARCH$(1,1)$ on the DAX and than GJR-GARCH on the
DAX and KOSPI. EGARCH is significantly more accurate on the S\&P~500 and Nikkei 225, and
FIGARCH is more accurate on KOSPI and Bitcoin; most remaining differences are not
significant. HAR-RV has the lowest QLIKE in all six markets, consistent with its direct
use of realized-variance lags. Relative to symmetric LM-GARCH, the asymmetric extension
produces a significant gain only on the DAX. The full Diebold--Mariano comparison is reported in Supplementary Table~\ref{supp:tab:oosdm}.
These results support forecasting comparability, and not general forecast dominance.

\FloatBarrier

\subsection{Robustness to the fixed memory exponent}
\label{sec:emp_robust}

At $p=2.0$, the conventional $\chi^2_1$ memory test rejects for the DAX, Nikkei~225,
KOSPI, and Bitcoin, but not for the S\&P~500 or FTSE~100. The corresponding
$LR_{\mathrm{mem}}$ statistics are $7.7$, $23.9$, $30.4$, and $26.3$ for the four
rejections; the DAX reference $p$-value is about $0.005$. Thus the fixed-$p$ check
continues to place offset asymmetry in a subset of markets. We point out here that the finite-sample bootstrap
classification, including the DAX non-rejection, refers to the baseline $p=1.2$
specification. Supplementary Table~\ref{supp:tab:p2mem} reports both fixed-exponent columns.


\section{Conclusion}
\label{sec:conclusion}

The symmetric long-memory GARCH model of \citet{lee2026} assigns positive and
negative return innovations a common power-law kernel and therefore cannot determine
whether volatility asymmetry arises from the magnitude of the response, its
finite-horizon persistence profile, or both. In this paper we extend that model by allowing the kernel injected after a
positive or negative innovation to differ in two respects. The injection amplitudes may
differ, producing the usual level asymmetry, and the kernel offsets may differ, allowing
the two innovations to generate different finite-horizon persistence profiles. Because
$p$ is common across signs, the memory channel does not imply different asymptotic
power-law exponents. These modifications give two separately testable channels of
asymmetry while preserving the finite-dimensional
Markov representation and exact nesting of the symmetric long memory GARCH model.

Applying the model to five equity indices and Bitcoin, we reject the joint null that
both channels are symmetric in each of the six series. This is an overall result,
carried primarily by the level channel, and does not imply that both restrictions are
rejected separately in every market. Negative innovations generate a larger immediate
increase in conditional variance than positive innovations throughout the sample. When the kernel
offsets are also allowed to differ, we find evidence of sign-dependent memory in the
Nikkei~225, KOSPI, and Bitcoin. The memory restriction is not rejected for the DAX at
the five-percent level (bootstrap $p=0.066$), and it is not separately identified for
the S\&P~500 and FTSE~100 because the positive injection branch is nearly inactive.
Thus, in the markets where the memory restriction is rejected, negative innovations
have the larger short-run effect, whereas positive innovations generate smaller but more
persistent volatility components over the horizons emphasized by the fitted kernels. The
conventional asymptotic memory-test ordering remains similar when the exponent is fixed
at $p=2.0$ rather than $p=1.2$. Bootstrap calibration is conducted for the baseline
$p=1.2$ specification; under both the Gaussian parametric and symmetrized-residual
schemes, the five-percent channel classifications are unchanged.

For interior parameter configurations in which both sign branches are active, we also
show that a joint Foster--Lyapunov condition yields positive Harris recurrence, a unique
invariant distribution, and an ergodic law of large numbers. The numerical diagnostic
is negative for five of the six fitted configurations. For the S\&P~500 and FTSE~100, the log-parameterized estimates remain strictly
positive and therefore satisfy the theorem's active-branch parameter restriction, but
they lie extremely close to the inactive-branch boundary. We consequently treat their
stability calculations as boundary-sensitive rather than leaning on them as strong
empirical applications of the theorem. The fitted DAX, KOSPI, and
Bitcoin interior configurations satisfy the sufficient condition, although KOSPI has
essentially no margin. The Nikkei~225 is not covered by the sufficient certificate at
the fitted configuration. Section~\ref{sec:emp_stab} shows that imposing the reported
certificate margin costs only $0.61$ log-likelihood points for the Nikkei. We note that because the
condition is sufficient rather than necessary, failure to obtain the certificate is
not evidence that the fitted process is unstable. Under the conditions of
Proposition~\ref{prop:boundary}, the comparative statics show how changes in the
level and memory channels shift the diagnostic stability frontier.

FIAPARCH records the highest in-sample quasi-log-likelihood and the lowest BIC in all
six series, so ALM-GARCH does not provide a general fit advantage over established
asymmetric long-memory models. Out of sample, ALM-GARCH is broadly comparable with the
standard return-based volatility models and does not deliver uniformly superior
forecasts. Except for the DAX, its forecasts are similar to and not significantly
different from those of the symmetric long-memory model. HAR-RV remains the most
accurate model for the realized-variance target.

ALM-GARCH instead provides a separately testable distinction between the level and
memory channels. The level restriction is rejected across all six series, whereas
memory-channel rejections are concentrated in the Nikkei~225, KOSPI, and Bitcoin.
Separate identification is limited when one shock branch is nearly inactive, as in the
S\&P~500 and FTSE~100, where the data cannot distinguish equality of the memory
offsets from weak identification caused by an inactive positive branch. Across the fitted series, negative shocks generate the larger
immediate volatility response, while positive shocks can generate smaller but more
persistent components when the memory channel is separately identified.

\section*{Funding}
This work was supported by the National Research Foundation of Korea (NRF) grant funded by
the Korea government (MSIT) (No. RS-2026-25469087).

\section*{Conflict of interest}
The authors declare no conflict of interest.

\clearpage
\bibliographystyle{chicago}
\bibliography{references}

\clearpage
\section*{Supplementary Appendix}
\addcontentsline{toc}{section}{Supplementary Appendix}
This appendix contains the detailed proofs, full parameter estimates, additional
simulations and sensitivity checks, optimizer diagnostics, and supporting forecast
comparisons referenced in the main article. Main-text equation, theorem, table, and
figure numbers retain original numbering; supplementary material carry an
``S'' prefix.

\setcounter{section}{0}
\setcounter{subsection}{0}
\setcounter{table}{0}
\setcounter{figure}{0}
\setcounter{equation}{0}
\renewcommand{\thesection}{S\arabic{section}}
\renewcommand{\thesubsection}{\thesection.\arabic{subsection}}
\renewcommand{\thetable}{S\arabic{table}}
\renewcommand{\thefigure}{S\arabic{figure}}
\renewcommand{\theequation}{S\arabic{equation}}


\section{Detailed proofs}
\label{app:proofs}

Throughout we write
$r=r(c)=(1+\tau/c)^{-1}$,
$\rho(c)=r(c)^p$,
$\sigma^2=\mu+x$,
$\eta=\sigma\varepsilon$,
$U=\varepsilon^2$,
$z=\eta^2=(\mu+x)U$.
For a shock of sign $s$, the injection scale is $\xi_s$ and the reset
offset is $\gamma_s$, with $(\xi_+,\gamma_+)$ on $\eta\ge0$ and
$(\xi_-,\gamma_-)$ on $\eta<0$; recall that
$\xi_-=\xi_+(1+\delta)$. We use the moment functionals
\[
m(\kappa)
=
\E\!\left[\frac{\kappa}{\kappa+U}\right],
\qquad
m'(\kappa)
=
\E\!\left[\frac{U}{(\kappa+U)^2}\right].
\]

\subsection{Proof of Lemma~\ref{lem:envelope}}

Fix a sign branch $s$. By \eqref{eq:alm_update}, conditional on that
branch,
\begin{equation}
	c_n
	=
	\frac{x r(c)^p+\xi_s z}
	{(x/c)r(c)^{p+1}+(\xi_s/\gamma_s)z},
	\qquad
	z=(\mu+x)U.
	\label{eq:cupdate}
\end{equation}
Define
\[
\kappa_s(x,c)
:=
\frac{\gamma_s x r(c)^{p+1}}
{\xi_s c(\mu+x)},
\qquad
\bar\kappa_s(c)
:=
\frac{\gamma_s r(c)^{p+1}}
{\xi_s c}.
\]
Then
\[
\kappa_s(x,c)
=
\frac{x}{\mu+x}\bar\kappa_s(c)
\le
\bar\kappa_s(c).
\]
Thus $\kappa_s(x,c)\uparrow\bar\kappa_s(c)$ as $x\to\infty$, and
$\bar\kappa_s(c)=\sup_{x\ge0}\kappa_s(x,c)$.

Subtracting $c$ in \eqref{eq:cupdate} and using
$c\{1-r(c)\}=\tau r(c)$ gives
\begin{equation}
	c_n-c
	=
	-(c-\gamma_s)
	+
	(c+\tau-\gamma_s)
	\frac{\kappa_s(x,c)}
	{\kappa_s(x,c)+U}.
	\label{eq:cident}
\end{equation}
Taking expectations over $U$ and the independent sign yields
\begin{equation}
	D_c(x,c)
	=
	\sum_s q_s
	\left[
	-(c-\gamma_s)
	+
	(c+\tau-\gamma_s)
	m\!\left(\kappa_s(x,c)\right)
	\right].
	\label{eq:mixdrift}
\end{equation}

We first prove part~(iii). The same branchwise comparison will then be
used to prove part~(i). By the definition of $S_\delta$,
\begin{equation}
	D_c(x,c)-S_\delta(c)
	=
	\sum_s q_s(c+\tau-\gamma_s)
	\left[
	m\!\left(\kappa_s(x,c)\right)
	-
	m\!\left(\bar\kappa_s(c)\right)
	\right].
	\label{eq:drift-envelope-difference}
\end{equation}
Moreover,
\[
\bar\kappa_s(c)-\kappa_s(x,c)
=
\frac{\mu}{\mu+x}\bar\kappa_s(c).
\]
Hence, for $U>0$,
\begin{align*}
	0
	&\le
	\frac{\bar\kappa_s(c)}
	{\bar\kappa_s(c)+U}
	-
	\frac{\kappa_s(x,c)}
	{\kappa_s(x,c)+U}
	\\
	&=
	\frac{
		\{\bar\kappa_s(c)-\kappa_s(x,c)\}U
	}{
		\{\bar\kappa_s(c)+U\}
		\{\kappa_s(x,c)+U\}
	}
	\\
	&=
	\frac{\mu}{\mu+x}
	\frac{\bar\kappa_s(c)U}
	{
		\{\bar\kappa_s(c)+U\}
		\{\kappa_s(x,c)+U\}
	}
	\\
	&\le
	\frac{\mu}{\mu+x}.
\end{align*}
Taking expectations and using the definition of $m$ gives
\[
0
\le
m\!\left(\bar\kappa_s(c)\right)
-
m\!\left(\kappa_s(x,c)\right)
\le
\frac{\mu}{\mu+x}.
\]
Consequently, for every compact set
$K\subset[\gamma_{\min},\infty)$,
\begin{align*}
	\sup_{c\in K}
	\left|D_c(x,c)-S_\delta(c)\right|
	&\le
	\frac{\mu}{\mu+x}
	\sum_s q_s
	\sup_{c\in K}|c+\tau-\gamma_s| \to 0 
	\quad \text{as } x\to\infty.
\end{align*}
This proves part~(iii). No sign restriction on
$c+\tau-\gamma_s$ is required for this compact-uniform convergence. In
particular, the result remains valid for compact carrier sets that
intersect the region
\[
c<\gamma_{\max}-\tau,
\]
whenever this region is nonempty, even though
$c+\tau-\gamma_s$ may then be negative for at least one branch.

We next prove part~(i). The preceding comparison shows that
\[
m\!\left(\kappa_s(x,c)\right)
\le
m\!\left(\bar\kappa_s(c)\right).
\]
If $c+\tau-\gamma_s\ge0$ for both sign branches, multiplying the
branchwise inequalities by $c+\tau-\gamma_s$ and summing over $s$ in
\eqref{eq:mixdrift} gives
\[
D_c(x,c)\le S_\delta(c).
\]
This proves part~(i).

We next prove part~(ii). For each sign branch $s$, the corresponding
term in \eqref{eq:mixdrift} satisfies
\[
-(c-\gamma_s)
+
(c+\tau-\gamma_s)m\!\left(\kappa_s(x,c)\right)
\le
\begin{cases}
	\tau,
	& c+\tau-\gamma_s\ge0,\\[3pt]
	\gamma_s-c,
	& c+\tau-\gamma_s<0.
\end{cases}
\]
Indeed, when $c+\tau-\gamma_s\ge0$, the bound follows from
$m(\kappa_s(x,c))\le1$, whereas when $c+\tau-\gamma_s<0$, it follows
from $m(\kappa_s(x,c))\ge0$. Since
\[
c\ge\gamma_{\min},
\qquad
\gamma_s\le\gamma_{\max},
\]
we have
\[
\gamma_s-c
\le
\gamma_{\max}-\gamma_{\min}.
\]
Averaging over the two sign branches therefore gives
\[
D_c(x,c)
\le
\max\{\tau,\gamma_{\max}-\gamma_{\min}\}
=
\bar S.
\]
This proves part~(ii).
The same branchwise argument, with $\bar\kappa_s(c)$ in place of
$\kappa_s(x,c)$, gives
\[
S_\delta(c)\le\bar S
\qquad
\text{for all }c\ge\gamma_{\min}.
\]

Finally,
\[
\bar\kappa_s(c)
=
\frac{\gamma_s r(c)^{p+1}}{\xi_s c}
\longrightarrow0
\qquad
\text{as }c\to\infty.
\]
Since $U>0$ almost surely,
\[
\frac{\bar\kappa_s(c)}
{\bar\kappa_s(c)+U}
\longrightarrow0
\qquad
\text{almost surely},
\]
and the ratio is bounded by one. Dominated convergence therefore gives
\[
m\!\left(\bar\kappa_s(c)\right)\longrightarrow0.
\]
It follows that
\[
(c+\tau-\gamma_s)
m\!\left(\bar\kappa_s(c)\right)
=
o(c),
\]
so each branch term in \eqref{eq:envelope} satisfies
\[
-(c-\gamma_s)
+
(c+\tau-\gamma_s)m\!\left(\bar\kappa_s(c)\right)
=
-(c-\gamma_s)+o(c).
\]
Consequently,
\[
S_\delta(c)\longrightarrow-\infty
\qquad
\text{as }c\to\infty.
\]
In the symmetric case, the two branch-specific terms are identical,
and $S_\delta$ reduces to the carrier-drift envelope of
\citet{lee2026}. \qed

\subsection{Proof of Proposition~\ref{prop:regularity}}

Let $P$ denote the one-step transition kernel of the chain and, for a measurable
function $f$, write
\[
Pf(x,c)=\E\!\left[f(X_n,c_n)\mid X_{n-1}=x,c_{n-1}=c\right].
\]
Fix one active branch $s\in\{+,-\}$. Under
Assumption~\ref{ass:innov}, the branch indicator has probability
$q_s>0$ and is independent of $U$. The random variable $U$ has a
lower-semicontinuous density $g$ that is strictly positive on
$(0,\infty)$. In particular, $g$ has a positive lower bound on every compact
interval contained in $(0,\infty)$.

Use $u=U$ as the control and define the branch map
\[
\Phi_s(x,c;u)
=
\left(
 x r(c)^p+\xi_s(\mu+x)u,
 \frac{x r(c)^p+\xi_s(\mu+x)u}
 {(x/c)r(c)^{p+1}+(\xi_s/\gamma_s)(\mu+x)u}
\right),
\qquad u>0.
\]

\paragraph{Attraction to the branch anchor.}
Let \( a_s=(0,\gamma_s).\)
For $x>0$, the continuous extension of $\Phi_s$ in the control
variable to $u=0$ satisfies
\begin{align*}
	\Phi_s(x,c;0)
	=
	\left(
	x r(c)^p,\,
	\frac{x r(c)^p}
	{(x/c)r(c)^{p+1}}
	\right) =
	\left(
	x r(c)^p,\,
	c+\tau
	\right).
\end{align*}
Iterating this limiting map $\ell$ times gives
\[
x_\ell
=
x\left(\frac{c}{c+\ell\tau}\right)^p
\longrightarrow0,
\qquad
c_\ell
=
c+\ell\tau
\longrightarrow\infty
\qquad
\text{as }\ell\to\infty.
\]

Define the decayed contribution from the previous state by
\[
\widetilde x_\ell
:=
x_\ell r(c_\ell)^p.
\]
Choose the final positive control as
\[
u_\ell
:=
\frac{\sqrt{\widetilde x_\ell}}
{\xi_s(\mu+x_\ell)},
\]
then the resulting contribution of the new shock to the $X$-coordinate is
\[
\xi_s(\mu+x_\ell)u_\ell
=
\sqrt{\widetilde x_\ell}.
\]
Since $\widetilde x_\ell\to0$ as $\ell \to \infty$,  the next $X$-coordinate satisfies
\[
x_{\ell+1}
=
\widetilde x_\ell+\sqrt{\widetilde x_\ell}
\longrightarrow0
\qquad
\text{as }\ell\to\infty.
\]

Moreover, taking the reciprocal of the $c$-update in
\eqref{eq:alm_update} gives
\[
\frac{1}{c_{\ell+1}}
=
\frac{\widetilde x_\ell}
{\widetilde x_\ell+\sqrt{\widetilde x_\ell}}
\frac{1}{c_\ell+\tau}
+
\frac{\sqrt{\widetilde x_\ell}}
{\widetilde x_\ell+\sqrt{\widetilde x_\ell}}
\frac{1}{\gamma_s}.
\]
Since
\[
\frac{\widetilde x_\ell}
{\widetilde x_\ell+\sqrt{\widetilde x_\ell}}
=
\frac{\sqrt{\widetilde x_\ell}}
{1+\sqrt{\widetilde x_\ell}}
\longrightarrow0, \qquad
\frac{\sqrt{\widetilde x_\ell}}
{\widetilde x_\ell+\sqrt{\widetilde x_\ell}}
=
\frac{1}
{1+\sqrt{\widetilde x_\ell}}
\longrightarrow1,
\]
as $\ell \to \infty$,
we obtain
\[
\frac{1}{c_{\ell+1}}
\longrightarrow
\frac{1}{\gamma_s}.
\]
Therefore,
\[
(x_{\ell+1},c_{\ell+1})
\longrightarrow
(0,\gamma_s)
=
a_s.
\]

If $x=0$, any positive control $u$ gives
\[
\Phi_s(0,c;u)
=
(\xi_s\mu u,\gamma_s),
\]
which can be made arbitrarily close to $a_s$ by choosing $u>0$
sufficiently small.

It remains to replace the zero controls used above by admissible
positive controls. Let $G$ be any neighborhood of $a_s$. For $x>0$,
the convergence established above implies that there exists a
sufficiently large but finite integer $L$ such that the terminal state
of the limiting path corresponding to $\ell=L$ lies in $G$.

By continuity of the finite-step composition in the control variables,
the first $L$ zero controls can be replaced by sufficiently small
positive controls, and the final control can be chosen sufficiently
close to $u_L$, while keeping the terminal state inside $G$. More
precisely, there exist compact intervals
\[
I_1,\ldots,I_{L+1}\Subset(0,\infty)
\]
such that every control vector in
\[
I_1\times\cdots\times I_{L+1}
\]
drives the initial state into $G$ along a path that remains in branch
$s$. The case $x=0$ is handled similarly using the single positive
control described above.

Since the density $g$ of $U$ is bounded away from zero on each $I_k$ and branch $s$
occurs with probability $q_s>0$, the corresponding finite
positive-control path has positive probability under the full signed
chain. Therefore, every state can reach any prescribed neighborhood of
$a_s$ with positive probability through a finite path that remains in
branch $s$.

\paragraph{A full-rank two-step map at the anchor.}
For an initial state $\zeta\in\mathcal S$, define the two-step branch-$s$
control map by
\[
\Phi_{s,\zeta}^{(2)}(u_1,u_2)
:=
\Phi_s\bigl(\Phi_s(\zeta;u_1);u_2\bigr).
\]

Starting from the anchor $a_s=(0,\gamma_s)$, apply $u_1>0$ and write
\[
y:=\xi_s\mu u_1,
\qquad
r_s:=r(\gamma_s),
\qquad
\rho_s:=\rho(\gamma_s)=r_s^p.
\]
Then the state after the first control is
\[
\Phi_s(a_s;u_1)=(y,\gamma_s).
\]
Hence
\[
\Phi_{s,a_s}^{(2)}(u_1,u_2)
=
\left(
y\rho_s+\xi_s(\mu+y)u_2,\,
\gamma_s
\frac{y\rho_s+\xi_s(\mu+y)u_2}
{y\rho_s r_s+\xi_s(\mu+y)u_2}
\right).
\]
Direct differentiation at $u_2=0$ gives
\begin{equation}
	\left|
	\det D_{(u_1,u_2)}\Phi_{s,a_s}^{(2)}(u_1,0)
	\right|
	=
	\frac{\gamma_s\mu\xi_s(1-r_s)(1+\xi_su_1)}
	{r_s^2u_1}
	>0.
	\label{eq:anchor-jacobian}
\end{equation}
For any fixed $u_1^*>0$, continuity therefore permits a sufficiently
small $u_2^*>0$ such that
\[
\det D_u\Phi_{s,a_s}^{(2)}(u^*)\ne0,
\qquad
u^*:=(u_1^*,u_2^*)\in(0,\infty)^2.
\]

Treating $\zeta$ as a parameter and $u=(u_1,u_2)$ as the control variable,
the parameterized inverse-function theorem, applied at $(a_s,u^*)$,
yields a relative neighborhood $N_0$ of $a_s$ in $\mathcal S$, a
relatively compact open neighborhood
\(
\mathcal U\Subset(0,\infty)^2
\)
of $u^*$, and an open ball
\(
B\Subset\operatorname{int}(\mathcal S)
\)
such that, for every $\zeta\in N_0$, the restriction
\(
\Phi_{s,\zeta}^{(2)}|_{\mathcal U}
\)
is a $C^1$ diffeomorphism onto its image and
\(
B\subset\Phi_{s,\zeta}^{(2)}(\mathcal U).
\)
After shrinking $N_0$ and $\mathcal U$ if necessary, continuity of the
Jacobian yields a finite constant $J_0<\infty$ such that
\[
\left|\det D_u\Phi_{s,\zeta}^{(2)}(u)\right|
\le J_0,
\qquad
\zeta\in N_0,\quad u\in\mathcal U.
\]

Let $g_0>0$ be a common lower bound for the density $g$ of $U$ on the
two coordinate projections of $\overline{\mathcal U}$. Restricting two
successive transitions to branch $s$ and to controls in $\mathcal U$,
the change-of-variables formula gives
\begin{equation}
	P^2(\zeta,A)
	\ge
	\frac{q_s^2g_0^2}{J_0}
	\operatorname{Leb}(A\cap B),
	\qquad
	\zeta\in N_0.
	\label{eq:anchor-minorization}
\end{equation}
Hence $N_0$ is a two-step small set.

\paragraph{Irreducibility.}
The attraction argument implies that, for each $\zeta\in\mathcal S$, there is an
integer $n(\zeta)$ with $P^{n(\zeta)}(\zeta,N_0)>0$. Define
$\phi(A)=\operatorname{Leb}(A\cap B)$. If $\phi(A)>0$, then
\[
P^{n(\zeta)+2}(\zeta,A)
\ge
P^{n(\zeta)}(\zeta,N_0)
\frac{q_s^2g_0^2}{J_0}\phi(A)>0.
\]
The chain is therefore $\phi$-irreducible.

\paragraph{Compact sets are petite.}
Let $K\subset\mathcal S$ be compact. For every $\zeta\in K$, continuity of
a finite positive-control path from $\zeta$ into $N_0$ yields a neighborhood
$O_\zeta$ of $\zeta$, an integer $n_\zeta\ge1$, and a constant $\alpha_\zeta>0$ such
that
\[
P^{n_\zeta}(\widetilde\zeta,N_0)\ge\alpha_\zeta,
\qquad
\widetilde\zeta\in O_\zeta.
\]

By compactness, there exist $\zeta_1,\ldots,\zeta_m\in K$ such that
\(
K\subset\bigcup_{i=1}^m O_{\zeta_i}.
\)
For notational simplicity, write
\[
O_i:=O_{\zeta_i},
\qquad
n_i:=n_{\zeta_i},
\qquad
\alpha_i:=\alpha_{\zeta_i}.
\]
Define
\[
\nu(A)
:=
\frac{q_s^2g_0^2}{J_0}
\operatorname{Leb}(A\cap B).
\]
Then, by the Chapman--Kolmogorov equation and
\eqref{eq:anchor-minorization}, for every $\widetilde\zeta\in O_i$,
\begin{align*}
	P^{n_i+2}(\widetilde\zeta,A)
	&=
	\int_{\mathcal S}
	P^2(w,A)\,
	P^{n_i}(\widetilde\zeta,\D w)
	\\
	&\ge
	\int_{N_0}
	P^2(w,A)\,
	P^{n_i}(\widetilde\zeta,\D w)
	\\
	&\ge
	P^{n_i}(\widetilde\zeta,N_0)\nu(A)
	\\
	&\ge
	\alpha_i\nu(A).
\end{align*}

Choose a probability distribution $b$ on $\mathbb Z_+$ such that
\[
b(n_i+2)>0,
\qquad
i=1,\ldots,m.
\]
For every $\widetilde\zeta\in K$, choose an index $i$ such that
$\widetilde\zeta\in O_i$. Then
\begin{align*}
	\sum_{n\ge0} b(n)P^n(\widetilde\zeta,A)
	\ge
	b(n_i+2)P^{n_i+2}(\widetilde\zeta,A) \ge
	b(n_i+2)\alpha_i\nu(A).
\end{align*}
Therefore, with
\[
\epsilon_K
:=
\min_{1\le i\le m}
b(n_i+2)\alpha_i
>0,
\]
we have
\[
\sum_{n\ge0} b(n)P^n(\widetilde\zeta,A)
\ge
\epsilon_K\nu(A),
\qquad
\widetilde\zeta\in K.
\]
Thus $K$ is petite. Since $K$ was arbitrary, every compact subset of
$\mathcal S$ is petite. \qed

\subsection{Proof of Theorem~\ref{thm:stationarity}}

By Proposition~\ref{prop:regularity}, the chain is
$\phi$-irreducible and every compact subset of $\mathcal S$ is petite.

Let
\[
V(x,c)=\log(1+x)+\lambda c,
\]
where $\lambda>0$ is supplied by Condition~\ref{cond:stability}, and define
\[
L_x(c)
=
\E\!\left[
\log(1+X_n)-\log(1+x)
\mid X_{n-1}=x,\ c_{n-1}=c
\right].
\]
Then
\[
PV(x,c)-V(x,c)
=
L_x(c)+\lambda D_c(x,c).
\]

We first establish a global upper bound for the log drift.
Conditional on branch $s\in\{+,-\}$,
\begin{equation}
\frac{1+X_n}{1+x}
=
A_x(c)+B_x\xi_sU, \qquad
A_x(c)
=
\frac{1+x\rho(c)}{1+x},
\qquad
B_x
=
\frac{\mu+x}{1+x}. \label{eq:AB}
\end{equation}
Since $0<\rho(c)\le1$,
\[
A_x(c)\le1,
\qquad
B_x\le b_\mu:=\max\{1,\mu\}.
\]
Therefore
\[
L_x(c)
\le
\bar L
:=
\sum_{s\in\{+,-\}}
q_s
\E\log(1+b_\mu\xi_sU)
<\infty
\]
for every $(x,c)\in\mathcal S$.
Finiteness follows from $\E U=1$, since
$\log(1+aU)\le aU$ for every $a>0$.

\paragraph{1. Large $c$ region.}
Because $S_\delta(c)\to-\infty$ as $c\to\infty$, choose
\[
c^\dagger
\ge
\max\{\gamma_{\min},\gamma_{\max}-\tau\}
\]
large enough that
\[
\bar L+\lambda S_\delta(c)
\le
-\epsilon_1,
\qquad
c\ge c^\dagger,
\]
for some $\epsilon_1>0$.

For $c\ge c^\dagger$, we have
\[
c+\tau-\gamma_s\ge0
\qquad
\text{for both }s\in\{+,-\}.
\]
Hence Lemma~\ref{lem:envelope} gives
\[
D_c(x,c)\le S_\delta(c).
\]
Consequently,
\[
PV(x,c)-V(x,c)
\le
\bar L+\lambda S_\delta(c)
\le
-\epsilon_1,
\qquad
c\ge c^\dagger,
\]
uniformly in $x$.

\paragraph{2. Bounded $c$ and large $X$ state.}
Set \( K_0=[\gamma_{\min},c^\dagger] \)
and
\[
\epsilon_2
:=
-\sup_{c\ge\gamma_{\min}}
\left\{
\Lambda_\delta(c)+\lambda S_\delta(c)
\right\}
>0.
\]
We show that the one-step drift converges uniformly to the joint
far-field drift on $K_0$. We have
\[
\sup_{c\in K_0}
|A_x(c)-\rho(c)|
\le
\frac{1}{1+x},
\qquad
|B_x-1|
=
\frac{|\mu-1|}{1+x}.
\]
Moreover, since $0<\rho(c)\le1$,
\[
A_x(c)-\rho(c)
=
\frac{1-\rho(c)}{1+x}
\ge0.
\]
Hence, for every $x\ge0$ and $c\in K_0$,
\[
A_x(c)\ge\rho(c)\ge\rho(\gamma_{\min})>0.
\]
The mean-value inequality for the logarithm,
together with $\E U=1$, therefore gives, for each branch $s$,
\[
\sup_{c\in K_0}
\left|
\E\log\{A_x(c)+B_x\xi_sU\}
-
\E\log\{\rho(c)+\xi_sU\}
\right|
\longrightarrow0.
\]
After mixing over the two branches,
\[
\sup_{c\in K_0}
|L_x(c)-\Lambda_\delta(c)|
\longrightarrow0.
\]
Lemma~\ref{lem:envelope} likewise gives
\[
\sup_{c\in K_0}
|D_c(x,c)-S_\delta(c)|
\longrightarrow0.
\]
It follows that
\[
\sup_{c\in K_0}
\left|
\{PV(x,c)-V(x,c)\}
-
\{\Lambda_\delta(c)+\lambda S_\delta(c)\}
\right|
\longrightarrow0.
\]

Hence there exists $M<\infty$ such that, whenever
$x\ge M$ and $c\in K_0$,
\[
PV(x,c)-V(x,c)
\le
\Lambda_\delta(c)+\lambda S_\delta(c)+\frac{\epsilon_2}{2}
\le
-\frac{\epsilon_2}{2}.
\]

\paragraph{3. The remaining compact region.}
Let
\[
C
=
[0,M]\times[\gamma_{\min},c^\dagger]
\]
and
\[
\epsilon_0
=
\min\left\{
\epsilon_1,\frac{\epsilon_2}{2}
\right\}.
\]
If $(x,c)\notin C$, then either $c>c^\dagger$, or
$c\le c^\dagger$ and $x>M$. The preceding two steps therefore imply
\[
PV(x,c)-V(x,c)
\le
-\epsilon_0,
\qquad
(x,c)\notin C.
\]

It remains to control the drift on $C$. For $(x,c)\in C$,
\[
\E[\log(1+X_n)]
=
\log(1+x)+L_x(c)
\le
\log(1+M)+\bar L.
\]
Moreover,
\[
c_n
\le
\max\{c+\tau,\gamma_s\}
\le
\max\{c^\dagger+\tau,\gamma_{\max}\}<\infty.
\]
Thus $PV$ is bounded above on $C$.
Since $V$ is also bounded on $C$,
\[
b_C :=
\max\left\{ 0,\,
\sup_{(x,c)\in C}
\left[
PV(x,c)-V(x,c)+\epsilon_0
\right]
\right\}
<\infty.
\]
Therefore
\[
PV \le V-\epsilon_0+b_C\mathbf 1_C.
\]

Define
\[
\widetilde V
=
1+\frac{V}{\epsilon_0}.
\]
Then $\widetilde V\ge1$, $\widetilde V$ is finite everywhere on
$\mathcal S$, and
\[
P\widetilde V \le \widetilde V-1 +  \frac{b_C}{\epsilon_0}\mathbf 1_C.
\]
Since the chain is $\phi$-irreducible and $C$ is petite,
\citet[Theorem~11.3.4]{meyn2009markov} implies that the chain is
positive Harris recurrent. It therefore admits a unique invariant
probability measure $\pi$, and the usual strong law holds for all
$\pi$-integrable functionals. \qed

\subsection{Proof of Proposition~\ref{prop:boundary}}

For parts (i) and (iii), consider the equal-reset case
$\gamma_+=\gamma_-=\gamma$, and write
\[
F(\xi,\delta)
:=
F(\xi,\delta;\gamma,\gamma).
\]
Let $c_\delta^*$ denote the first zero of the corresponding memory-scale drift
envelope $S_\delta$, suppressing its dependence on $\xi$ and $\gamma$, so that
\[
S_\delta(c_\delta^*)=0,
\qquad
S_\delta'(c_\delta^*)<0.
\]
Recall that
\[
\rho(c)=r(c)^p,
\qquad
\rho'(c)
=
\rho(c)\frac{p\tau}{c(c+\tau)}
>0.
\]

\paragraph{(i) Level channel.}
We first consider the local effect at the symmetric null. Define
\[
F_0(\xi):=F(\xi,0).
\]
At $\delta=0$, the two branches coincide. Since varying $\delta$ at fixed
$\xi$ perturbs only the negative-shock amplitude,
$\xi_-=\xi(1+\delta)$, the negative branch contributes the fraction $q_-$ of
the symmetric first variation, with
$\partial\xi_-/\partial\delta=\xi$. Hence
\begin{equation}
	\left.
	\frac{\partial F(\xi,\delta)}{\partial\delta}
	\right|_{\delta=0}
	=
	q_-\,\xi\,\frac{\D}{\D\xi}F_0(\xi).
	\label{eq:localnull-pf}
\end{equation}
Therefore, if
\[
\frac{\D}{\D\xi}F_0(\xi)>0,
\]
a small positive level asymmetry increases $F$ and is locally destabilizing
around the symmetric frontier.

We next consider the global effect. At fixed $\xi$, changing $\delta$ affects
$F$ both directly through the negative-shock amplitude and indirectly through
the induced movement of $c_\delta^*$. Thus
\begin{equation}
	\frac{\partial F(\xi,\delta)}{\partial\delta}
	=
	q_-\xi\,
	\E\!\left[
	\frac{U}
	{\rho(c_\delta^*)+\xi(1+\delta)U}
	\right]
	+
\rho'(c_\delta^*)\,
	\frac{\partial c_\delta^*}{\partial\delta}\,
	W(c_\delta^*),
	\label{eq:Fdelta}
\end{equation}
where
\[
W(c_\delta^*)
=
q_+
\E\!\left[
\frac{1}{\rho(c_\delta^*)+\xi U}
\right]
+
q_-
\E\!\left[
\frac{1}
{\rho(c_\delta^*)+\xi(1+\delta)U}
\right].
\]

Since $c_\delta^*$ is defined implicitly by
$S_\delta(c_\delta^*)=0$, the implicit function theorem gives
\[
\frac{\partial c_\delta^*}{\partial\delta}
=
-
\frac{
	\left.
	\partial S_\delta(c)/\partial\delta
	\right|_{c=c_\delta^*}
}{
	S_\delta'(c_\delta^*)
}.
\]
Only the negative branch varies with $\delta$, and direct differentiation
yields
\[
\frac{\partial c_\delta^*}{\partial\delta}
=
\frac{
	q_-\,(c_\delta^*+\tau-\gamma)\,\kappa_-\,m'(\kappa_-)
}{
	(1+\delta)\,S_\delta'(c_\delta^*)
},
\]
Substituting this expression into \eqref{eq:Fdelta} gives
\[
\frac{\partial F(\xi,\delta)}{\partial\delta}
=
q_-\xi
\left\{
\E\!\left[
\frac{U}
{\rho(c_\delta^*)+\xi(1+\delta)U}
\right]
-
\frac{
\rho'(c_\delta^*)\,
	W(c_\delta^*)\,
	c_\delta^*+\tau-\gamma,
	\kappa_-\,m'(\kappa_-)
}{
	\xi(1+\delta)\,
	|S_\delta'(c_\delta^*)|
}
\right\}.
\]
Condition~\ref{cond:dominance} therefore implies
\[
\frac{\partial F(\xi,\delta)}{\partial\delta}\ge0
\]
over any range of $\delta$ on which the condition holds. Since $F$ is
increasing in the injection scale at the relevant frontier,
$\xi^*(\gamma,\gamma,\delta)$ is nonincreasing in $\delta$.

\paragraph{(ii) Memory channel.}
Set $\delta=0$ and consider a single branch
\[
T(c;\gamma)
=
\gamma-c+(c+\tau-\gamma)m\!\left(\bar\kappa(c;\gamma)\right),
\qquad
\bar\kappa(c;\gamma)
=
\frac{\gamma r(c)^{p+1}}{\xi c}.
\]
For brevity, write $\bar\kappa=\bar\kappa(c;\gamma)$. Differentiating with respect to
$\gamma$ gives
\[
\frac{\partial T}{\partial\gamma}
=
1-m(\bar\kappa)
+
(c+\tau-\gamma)m'(\bar\kappa)\frac{\bar\kappa}{\gamma}.
\]
This derivative is positive whenever $c+\tau-\gamma\ge0$. Therefore, if
\[
c_0^*+\tau-\gamma_-\ge0,
\]
lowering $\gamma_-$ lowers the negative-branch contribution to the
memory-scale drift envelope and hence lowers $S_0$ at the relevant crossing.
Since the crossing is simple and downward, its location $c_0^*$ decreases.
Because $\rho'(c)>0$, the corresponding
\[
\rho_0^*=\rho(c_0^*)
\]
also decreases, and hence
\[
F(\xi,0;\gamma_+,\gamma_-)
\]
decreases at a given injection scale $\xi$.

Since the diagnostic is increasing in $\xi$ at the frontier, a lower value of
$F$ allows a larger injection scale before the frontier is reached. Therefore,
\[
\xi^*(\gamma_+,\gamma_-,0)
\ge
\xi_0^*.
\]
Thus, a harder negative-shock reset is stabilizing.

\paragraph{(iii) Two-sided bracket.}
Return to the equal-reset case $\gamma_-=\gamma_+$. By assumption,
\[
F(\xi_0^*,\delta;\gamma_+,\gamma_+)\ge0
\]
and
\[
F\!\left(
\frac{\xi_0^*}{1+\delta},
\delta;
\gamma_+,\gamma_+
\right)\le0.
\]
Thus, the diagnostic is nonpositive at
$\xi_0^*/(1+\delta)$ and nonnegative at $\xi_0^*$. By continuity and the
assumed uniqueness of the relevant diagnostic frontier, it follows that
\[
\frac{\xi_0^*}{1+\delta}
\le
\xi^*(\gamma_+,\gamma_+,\delta)
\le
\xi_0^*.
\]
\qed

\subsection{Note on the symmetric null}

At $(\xi_-,\gamma_-)=(\xi_+,\gamma_+)$ each result reduces to its base-model counterpart:
$S_\delta\to S$, $\Lambda_\delta\to\Lambda$, $F\to F_0$, Condition~\ref{cond:stability}
becomes the joint stability assumption of \citet{lee2026}, and the control model becomes
the one-regime model. The likelihood-ratio tests of Section~\ref{sec:identification} are
therefore exact tests of a nested restriction.

\section{Full parameter estimates}
\label{supp:estimates}

Table~\ref{tab:almest} reports the complete ALM-GARCH parameter estimates at fixed
$p=1.2$ underlying the channel tests of Section~\ref{sec:empirics}.

\begin{table}[H]
\centering
\caption{ALM-GARCH parameter estimates at fixed $p=1.2$.}
\label{tab:almest}
\begin{tabular}{lrrrrrrr}
\toprule
Asset & $\hat\mu$ & $\hat\xi_+$ & $\hat\xi_-$ & $\hat\rho_+$ & $\hat\rho_-$ & $\hat\gamma_+$ & $\log L$ \\
\midrule
S\&P 500 & $1.24\times10^{-5}$ & $1.654\times10^{-5}$ & 0.201 & 0.977 & 0.801 & 50.1 & 19694.0 \\
FTSE 100 & $1.55\times10^{-5}$ & $4.592\times10^{-7}$ & 0.202 & 0.986 & 0.793 & 85.2 & 19561.8 \\
DAX & $1.76\times10^{-5}$ & $1.238\times10^{-3}$ & 0.151 & 0.999 & 0.839 & 1551.0 & 19026.6 \\
Nikkei 225 & $1.50\times10^{-5}$ & 0.0471 & 0.179 & 0.999 & 0.733 & 1887.9 & 18569.8 \\
KOSPI & $7.94\times10^{-6}$ & 0.0299 & 0.129 & 0.998 & 0.813 & 556.6 & 18959.8 \\
Bitcoin & $3.33\times10^{-4}$ & 0.1173 & 0.259 & 0.875 & 0.528 & 8.5 & 13657.4 \\
\bottomrule
\end{tabular}

\tabnote{$\hat\mu$ is the variance intercept; $\hat\xi_\pm$ are the positive- and
negative-shock injection amplitudes; $\hat\rho_\pm$ are the corresponding persistence
summaries; $\hat\gamma_+$ is the positive-branch offset; and $\log L$ is the maximized
Gaussian quasi-log-likelihood.}
\end{table}

\section{Sandwich-robust standard errors}
\label{supp:qml_se}

Table~\ref{tab:qml_se} reports quasi-maximum-likelihood standard errors for the channel
quantities, computed at the estimates of Table~\ref{tab:almest} with the sandwich
covariance $n^{-1}H^{-1}JH^{-1}$, where $H$ is the negative average Hessian and $J$ the average
outer product of per-observation scores of the Gaussian quasi-likelihood, both obtained
by numerical differentiation in the estimation parameterization, with the delta method
applied to $(\hat\xi_+,\hat\xi_-,\hat\rho_+,\hat\rho_-)$. For the S\&P~500 and
FTSE~100 the fitted positive amplitude lies below the pre-specified tolerance
$\xi_{\mathrm{tol}}=10^{-4}$: $\hat\xi_+$ then sits at the parameter boundary and
$\hat\rho_+$ is not separately identified, so no covariance-based standard error is
meaningful for these cells and they are omitted (daggers). These values are descriptive;
the channel conclusions in the main text rest on the likelihood-ratio tests with
bootstrap calibration, not on Wald inference.

\begin{table}[H]
\centering
\caption{Sandwich-robust QML standard errors at fixed $p=1.2$.}
\label{tab:qml_se}
\begin{tabular}{lcccc}
\toprule
Asset & $\hat\xi_+$ (s.e.) & $\hat\xi_-$ (s.e.) & $\hat\rho_+$ (s.e.) & $\hat\rho_-$ (s.e.) \\
\midrule
S\&P 500 & ---\,$^{\dagger}$ & 0.201 (0.077) & ---\,$^{\dagger}$ & 0.801 (0.034) \\
FTSE 100 & ---\,$^{\dagger}$ & 0.202 (0.023) & ---\,$^{\dagger}$ & 0.793 (0.020) \\
DAX & $1.24\times10^{-3}$ ($4.92\times10^{-3}$) & 0.151 (0.021) & 0.999 ($1.71\times10^{-4}$) & 0.839 (0.019) \\
Nikkei 225 & 0.047 (0.013) & 0.179 (0.048) & 0.999 ($3.75\times10^{-4}$) & 0.733 (0.045) \\
KOSPI & 0.030 ($5.42\times10^{-3}$) & 0.129 (0.022) & 0.998 ($6.42\times10^{-4}$) & 0.813 (0.025) \\
Bitcoin & 0.117 (0.027) & 0.259 (0.062) & 0.875 (0.077) & 0.528 (0.060) \\
\bottomrule
\end{tabular}

\tabnote{Entries are point estimates with sandwich standard errors in parentheses.
$^{\dagger}$ denotes a fitted positive amplitude below the pre-specified tolerance
$\xi_{\mathrm{tol}}=10^{-4}$; the associated positive-branch quantity is at the boundary or
not separately identified, so a covariance-based standard error is not reported.}
\end{table}


\section{Data construction and fixed-\texorpdfstring{$p$}{p} diagnostics}

\begin{figure}[hbt!]
\centering
\includegraphics[width=\linewidth]{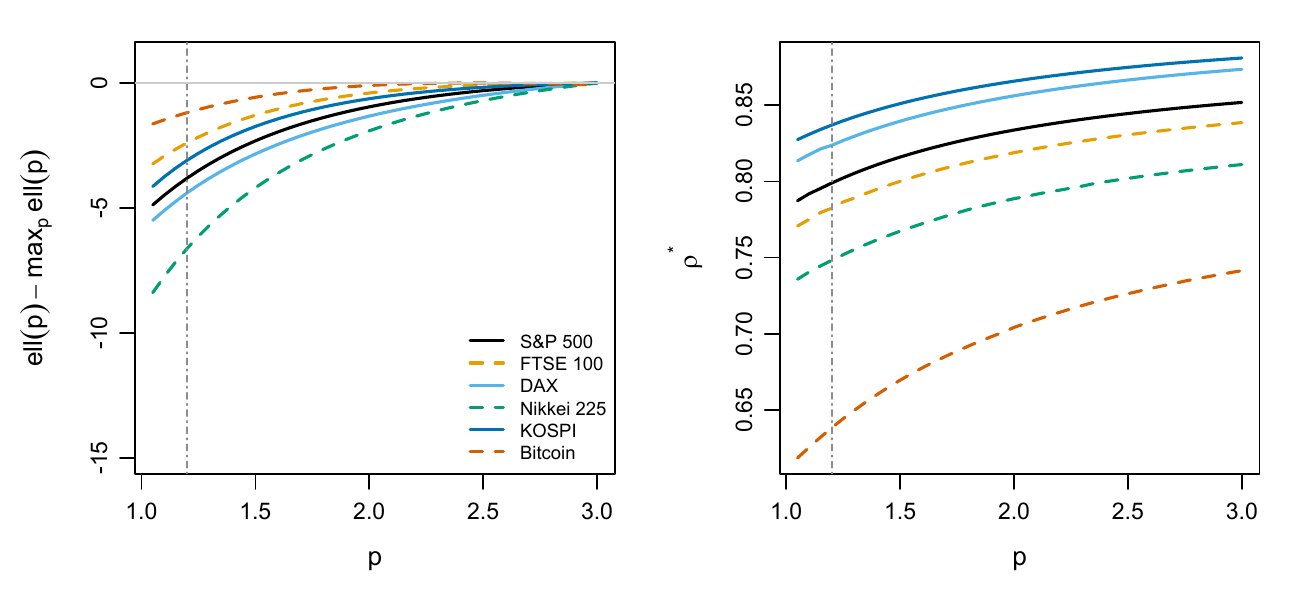}
\caption{Profile evidence on the fixed memory exponent. The left panel reports the
profile log-likelihood centered at each asset's maximum; the right panel reports the
implied persistence summary.}
\label{supp:fig:profile}
\end{figure}

\begin{figure}[hbt!]
\centering
\includegraphics[width=0.85\linewidth]{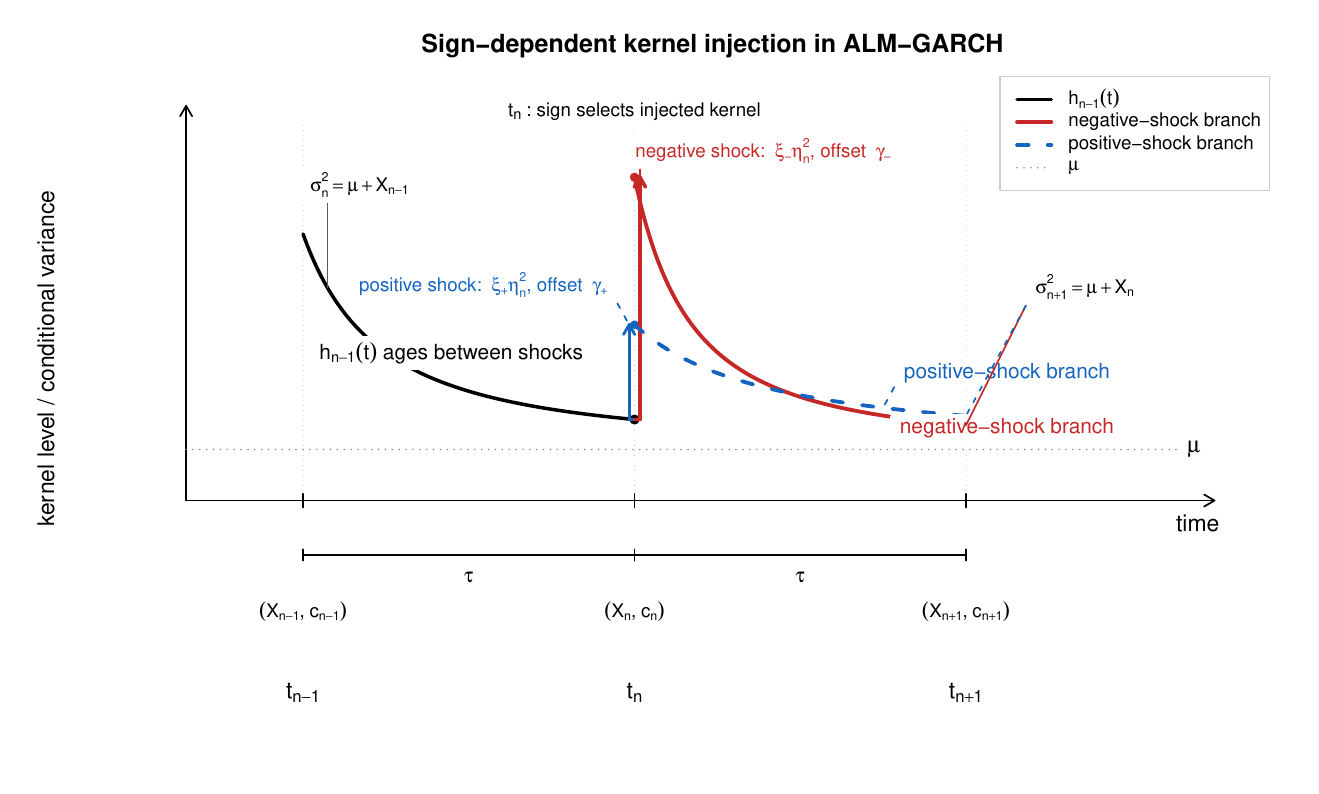}
\caption{Sign-dependent kernel update. Between arrivals the existing kernel ages
deterministically. At an arrival, shock sign selects a branch-specific injection
amplitude and reset offset.}
\label{supp:fig:kernel}
\end{figure}

Table~\ref{supp:tab:data} gives the exact samples and distinguishes the realized-variance
frequency used for Bitcoin from that used for the equity indices. Figure~\ref{supp:fig:profile}
and Table~\ref{supp:tab:profile} then report the profile-likelihood and persistence
sensitivity to the fixed memory exponent, while Figure~\ref{supp:fig:kernel} illustrates
the sign-dependent kernel update.

\begin{table}[hbt!]
\centering
\footnotesize
\caption{Data sources and sample construction.}
\label{supp:tab:data}
\begin{tabular}{@{}llrl>{\raggedright\arraybackslash}p{5.2cm}@{}}
\toprule
Asset & Sample period & $N$ & Source & Daily series \\
\midrule
S\&P 500   & 2000-01-03--2018-06-27 & 4641 & Oxford--Man &
Open-to-close return and five-minute realized variance \\
FTSE 100   & 2000-01-04--2018-06-27 & 4660 & Oxford--Man &
Open-to-close return and five-minute realized variance \\
DAX        & 2000-01-03--2018-06-27 & 4692 & Oxford--Man &
Open-to-close return and five-minute realized variance \\
Nikkei 225 & 2000-02-02--2018-06-27 & 4501 & Oxford--Man &
Open-to-close return and five-minute realized variance \\
KOSPI      & 2000-01-04--2018-06-27 & 4550 & Oxford--Man &
Open-to-close return and five-minute realized variance \\
Bitcoin    & 2012-01-02--2025-01-06 & 4754 & Bitstamp &
UTC close-to-close return and one-minute realized variance \\
\bottomrule
\end{tabular}

\tabnote{The equity series are from the Oxford--Man Institute Realized Library
\citep{heber2009}. Bitcoin prices are aligned to a complete one-minute UTC grid. Missing
minutes carry forward the previous close and therefore contribute a zero one-minute
return. Daily Bitcoin realized variance is the sum of the 1,439 squared one-minute log
returns within each UTC day.}
\end{table}

\begin{table}[hbt!]
\centering
\caption{Sensitivity to the fixed memory exponent.}
\label{supp:tab:profile}
\begin{tabular}{lrrrrr}
\toprule
\multicolumn{6}{l}{\emph{Log-likelihood cost} $\ell(p)-\max_{p}\ell(p)$} \\
\midrule
Asset & $p{=}1.1$ & $p{=}1.2$ & $p{=}1.5$ & $p{=}2.0$ & $p{=}3.0$ \\
\midrule
S\&P 500 & -4.5 & -3.8 & -2.3 & -1.0 & 0.0 \\
FTSE 100 & -2.9 & -2.4 & -1.3 & -0.4 & 0.0 \\
DAX & -5.1 & -4.4 & -2.8 & -1.3 & 0.0 \\
Nikkei 225 & -7.7 & -6.6 & -4.2 & -1.9 & 0.0 \\
KOSPI & -3.8 & -3.1 & -1.7 & -0.6 & 0.0 \\
Bitcoin & -1.5 & -1.2 & -0.6 & -0.1 & -0.1 \\
\midrule
\multicolumn{6}{l}{\emph{Implied persistence} $\rho$} \\
\midrule
Asset & $p{=}1.1$ & $p{=}1.2$ & $p{=}1.5$ & $p{=}2.0$ & $p{=}3.0$ \\
\midrule
S\&P 500 & 0.792 & 0.799 & 0.816 & 0.834 & 0.852 \\
FTSE 100 & 0.775 & 0.782 & 0.800 & 0.819 & 0.839 \\
DAX & 0.818 & 0.824 & 0.840 & 0.856 & 0.874 \\
Nikkei 225 & 0.740 & 0.748 & 0.767 & 0.789 & 0.811 \\
KOSPI & 0.831 & 0.837 & 0.851 & 0.866 & 0.881 \\
Bitcoin & 0.625 & 0.638 & 0.670 & 0.704 & 0.741 \\
\bottomrule
\end{tabular}

\tabnote{The upper block reports the profile log-likelihood at each fixed exponent,
centered at the largest value within each asset. The lower block reports the implied
persistence summary $\rho$.}
\end{table}

\section{Boundary sensitivity and cross-market channel summary}

\begin{figure}[hbt!]
\centering
\includegraphics[width=\linewidth]{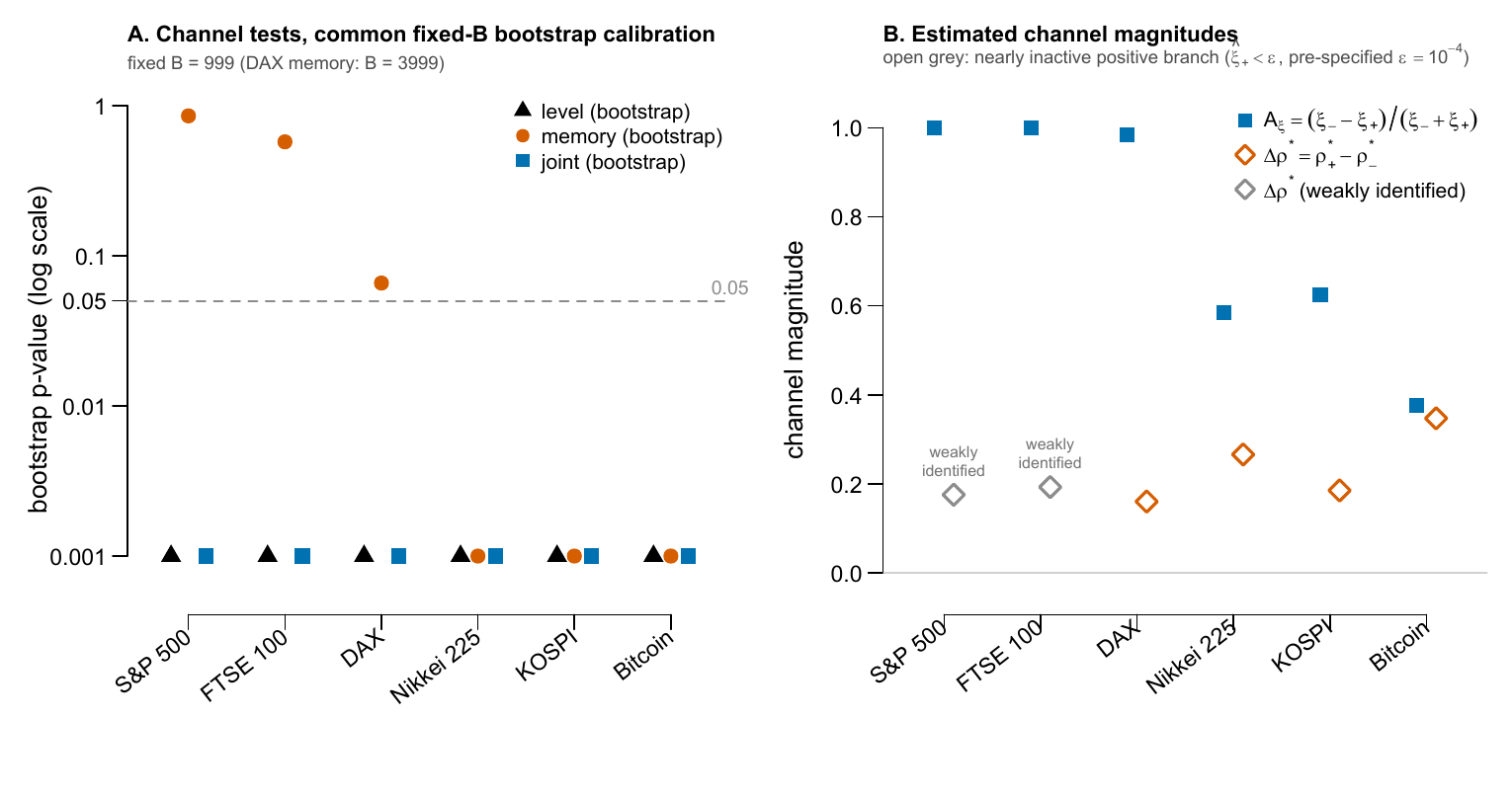}
\caption{Cross-market evidence on the two asymmetry channels at fixed $p=1.2$.
Panel~A reports fixed-$B$ parametric-bootstrap $p$-values; Panel~B reports normalized
level-channel asymmetry and the persistence difference $\rho_+-\rho_-$.}
\label{supp:fig:crossmarket}
\end{figure}

Table~\ref{supp:tab:floor} reports the floor sensitivity of the memory-channel test,
and Figure~\ref{supp:fig:crossmarket} summarizes the cross-market evidence for the two
asymmetry channels.

\begin{table}[hbt!]
\centering
\caption{Floor sensitivity of the memory-channel likelihood-ratio statistic at
fixed $p=1.2$.}
\label{supp:tab:floor}
\begin{tabular}{lrrrr}
\toprule
Asset & $\underline\xi_+=0$ & $\underline\xi_+=0.01$ & $\underline\xi_+=0.02$ & $\underline\xi_+=0.05$ \\
\midrule
S\&P 500 & 0.15 & 0.41 & 0.25 & 0.06 \\
FTSE 100 & 0.94 & 1.96 & 1.97 & 1.72 \\
DAX & 7.23$^{***}$ & 7.54$^{***}$ & 7.05$^{***}$ & 6.00$^{**}$ \\
Nikkei 225 & 15.01$^{***}$ & 15.01$^{***}$ & 15.01$^{***}$ & 15.01$^{***}$ \\
KOSPI & 26.08$^{***}$ & 26.08$^{***}$ & 26.08$^{***}$ & 25.57$^{***}$ \\
Bitcoin & 24.82$^{***}$ & 24.82$^{***}$ & 24.82$^{***}$ & 24.82$^{***}$ \\
\bottomrule
\end{tabular}

\tabnote{Entries are $LR_{\mathrm{mem}}$ when the positive injection amplitude is
constrained to lie at or above the displayed floor $\underline\xi_+$. Each specification
is estimated by the same multi-start procedure. Stars use the conventional $\chi^2_1$
reference: $^{*}p<0.10$, $^{**}p<0.05$, and $^{***}p<0.01$.}
\end{table}

\subsection{Robustness of the memory test to the fixed exponent}

Table~\ref{supp:tab:p2mem} reports the multistart-corrected memory-channel statistic at
$p=1.2$ and $p=2.0$. The $p$-values in this table use the conventional $\chi^2_1$
reference and are included only as a specification check; the finite-sample bootstrap
calibration in the main article is conducted for the baseline $p=1.2$ specification.

\begin{table}[hbt!]
\centering
\caption{Memory-channel likelihood-ratio statistic at fixed $p=1.2$ and $p=2.0$.}
\label{supp:tab:p2mem}
\begin{tabular}{lrrrr}
\toprule
Asset & $LR_{\mathrm{mem}}^{1.2}$ & $p_{\chi^2}^{1.2}$ & $LR_{\mathrm{mem}}^{2.0}$ & $p_{\chi^2}^{2.0}$ \\
\midrule
S\&P 500   & 0.15 & 0.697 & 0.18 & 0.673 \\
FTSE 100   & 0.95 & 0.330 & 0.82 & 0.365 \\
DAX        & 7.23 & 0.007 & 7.72 & 0.005 \\
Nikkei 225 & 15.01 & $<0.001$ & 23.90 & $<0.001$ \\
KOSPI      & 26.08 & $<0.001$ & 30.37 & $<0.001$ \\
Bitcoin    & 24.82 & $<0.001$ & 26.28 & $<0.001$ \\
\bottomrule
\end{tabular}
\tabnote{$LR_{\mathrm{mem}}$ tests $\gamma_-=\gamma_+$. The $p$-values use the
conventional $\chi^2_1$ reference, are reported to three decimals, and are shown as
$p<0.001$ when smaller than 0.001. The $p=2.0$ columns use the best unrestricted and
memory-restricted likelihoods from the multi-start convergence diagnostic. These
asymptotic values are reference calibrations; Table~\ref{tab:boot} of the main article
gives the baseline finite-sample bootstrap inference.}
\end{table}

\section{Monte Carlo design and finite-sample calibration}

The representative and fitted-configuration exercises serve distinct purposes. The
representative design uses $1{,}000$ samples of length $4{,}000$ under an asymmetric
configuration and an interior symmetric null; Tables~\ref{supp:tab:mcacc} and
\ref{supp:tab:mcsize} report estimator accuracy and channel-test size and power.
Table~\ref{supp:tab:nullsize} instead reuses the fixed-$B$ bootstrap replications to
measure the size of the conventional asymptotic tests at the fitted restricted nulls.
For the memory restriction, the S\&P~500, FTSE~100, and DAX null configurations are
boundary-prone, even though the unrestricted DAX fit has an active positive branch.
Finally, Table~\ref{supp:tab:mccalib} reports a separate unrestricted fitted-design
experiment for the two empirically difficult near-boundary fits, the S\&P~500 and
FTSE~100, using $500$ samples per design at the empirical sample length.

\begin{table}[hbt!]
\centering
\caption{Monte Carlo accuracy of the quasi-maximum-likelihood estimator at fixed $p=1.2$.}
\label{supp:tab:mcacc}
\begin{tabular}{lrr}
\toprule
Parameter & Bias & RMSE \\
\midrule
$\mu$ & 0.000000 & 0.000002 \\
$\xi$ & -0.0001 & 0.0182 \\
$\rho_+$ & 0.0011 & 0.0435 \\
$\rho_-$ & -0.0014 & 0.0347 \\
$\delta$ & 0.0522 & 0.3673 \\
\bottomrule
\end{tabular}

\tabnote{Results use $1{,}000$ samples of length $N=4{,}000$ generated under the
asymmetric configuration. Bias is the mean estimation error and RMSE is the root
mean-squared error across replications.}
\end{table}

\begin{table}[hbt!]
\centering
\caption{Monte Carlo size and power of the channel tests at fixed $p=1.2$.}
\label{supp:tab:mcsize}
\begin{tabular}{lrr}
\toprule
Test & Size & Power \\
\midrule
$LR_{\mathrm{lev}}$ & 0.060 & 0.302 \\
$LR_{\mathrm{mem}}$ & 0.053 & 0.199 \\
$LR_{\mathrm{sym}}$ & 0.056 & 0.226 \\
\bottomrule
\end{tabular}

\tabnote{Results use $1{,}000$ samples of length $N=4{,}000$. Size is the rejection
frequency under the symmetric null and power is the rejection frequency under the
asymmetric data-generating process; the nominal level is 0.05.}
\end{table}

The null score correlation between the level and memory directions is
0.95
\unskip, explaining the low power of separate channel
tests close to symmetry.

\begin{table}[hbt!]
\centering
\caption{Empirical size of the asymptotic channel tests at the fitted restricted nulls.}
\label{supp:tab:nullsize}
\begin{tabular}{lccc}
\toprule
Asset & level & memory & joint \\
\midrule
S\&P 500 & 0.047 (0.007) & 0.246 (0.014) & 0.045 (0.007) \\
FTSE 100 & 0.041 (0.006) & 0.197 (0.013) & 0.053 (0.007) \\
DAX & 0.052 (0.007) & 0.220 (0.007) & 0.050 (0.007) \\
Nikkei 225 & 0.042 (0.006) & 0.057 (0.007) & 0.055 (0.007) \\
KOSPI & 0.066 (0.008) & 0.056 (0.007) & 0.050 (0.007) \\
Bitcoin & 0.059 (0.007) & 0.058 (0.007) & 0.053 (0.007) \\
\bottomrule
\end{tabular}

\tabnote{Entries are rejection frequencies at the nominal five-percent level, with
Monte Carlo standard errors in parentheses, extracted from the Gaussian fixed-$B$
bootstrap replications. The level and memory tests use one degree of freedom and the
joint test uses two. $B=999$ for each cell except the DAX memory null, for which
$B=3{,}999$.}
\end{table}

\begin{table}[hbt!]
\centering
\caption{Estimation at empirically difficult fitted configurations.}
\label{supp:tab:mccalib}
\begin{tabular}{lrrrrrr}
\toprule
Design & bhit$_+$ & \multicolumn{2}{c}{$\hat\rho_+$ (uncond.)} & \multicolumn{2}{c}{$\hat\rho_+$ (cond.)} & RMSE $\hat\rho_-$ \\
\cmidrule(lr){3-4}\cmidrule(lr){5-6}
 & freq. & median & IQR & median & IQR & \\
\midrule
S\&P 500 & 0.48 & 0.70 & [0.43, 0.95] & 0.72 & [0.55, 0.93] & 0.018 \\
FTSE 100 & 0.45 & 0.71 & [0.44, 0.94] & 0.70 & [0.56, 0.89] & 0.020 \\
\bottomrule
\end{tabular}

\tabnote{Each design uses $500$ samples at the corresponding empirical sample length.
$\mathrm{bhit}_+$ is the frequency with which $\hat\xi_+<10^{-4}$. Unconditional
summaries use all converged replications; conditional summaries exclude replications in
which the positive branch is pinned to that boundary.}
\end{table}

\section{Optimization and convergence diagnostics}

The memory-channel likelihood-ratio statistic is based on multi-start estimation of
both unrestricted and restricted models. Table~\ref{tab:lrmem-convergence-p12} reports
the best likelihoods and repeated near-best solutions; Table~\ref{tab:lrmem-spread-audit}
reports the spread of the restricted fits at both fixed exponents.

\begin{table}[!htbp]
\centering
\caption{Multi-start convergence check for the ALM memory-channel LR test at fixed $p=1.2$.}
\label{tab:lrmem-convergence-p12}
\begin{tabular}{lrrrrrl}
\toprule
Asset & $\ell_U$ & $\ell_M$ & $LR_{\mathrm{mem}}$ & $p$-value & Near-best & Flag \\
\midrule
Bitcoin & 13657.45 & 13645.04 & 24.82 & $<0.001$ & 49/66 & ok \\
DAX & 19026.59 & 19022.98 & 7.23 & 0.007 & 8/11 & ok \\
FTSE 100 & 19561.79 & 19561.32 & 0.95 & 0.330 & 8/13 & ok \\
KOSPI & 18959.84 & 18946.80 & 26.08 & $<0.001$ & 11/58 & ok \\
Nikkei 225 & 18569.81 & 18562.30 & 15.01 & $<0.001$ & 4/63 & ok \\
S\&P 500 & 19694.02 & 19693.94 & 0.15 & 0.697 & 4/8 & ok \\
\bottomrule
\end{tabular}
\tabnote{The unrestricted ALM and memory-restricted model ($\gamma_-=\gamma_+$)
are refitted from multiple starting values. Near-best counts report the number of
successful unrestricted/restricted fits within the log-likelihood tolerance of the best
fit. $LR_{\mathrm{mem}}$ uses the best unrestricted and restricted log-likelihoods. The
$p$-value uses the conventional $\chi^2_1$ reference; values below 0.001 are reported as
$p<0.001$.}
\end{table}

\begin{table}[!htbp]
\centering
\caption{Likelihood-spread audit for the memory-restricted ALM multi-start fits.}
\label{tab:lrmem-spread-audit}
\begin{tabular}{llrrrrr}
\toprule
Asset & $p$ & Initial ok & Near-best & Best--2nd & Best--5th & Full range \\
\midrule
Bitcoin & 1.2 & 64 & 66 & 0 & $7.28\times10^{-12}$ & 1043.50 \\
DAX & 1.2 & 64 & 11 & $2.68\times10^{-7}$ & $1.63\times10^{-4}$ & 830.84 \\
FTSE 100 & 1.2 & 64 & 13 & 0 & $1.39\times10^{-4}$ & 1074.28 \\
KOSPI & 1.2 & 64 & 58 & 0 & $1.46\times10^{-11}$ & 961.12 \\
Nikkei 225 & 1.2 & 64 & 63 & $3.64\times10^{-12}$ & $1.09\times10^{-11}$ & 703.65 \\
S\&P 500 & 1.2 & 64 & 8 & $7.87\times10^{-9}$ & $4.51\times10^{-4}$ & 1235.73 \\
Bitcoin & 2.0 & 64 & 64 & 0 & $2.91\times10^{-11}$ & 1044.46 \\
DAX & 2.0 & 64 & 2 & $2.63\times10^{-6}$ & $1.66\times10^{-3}$ & 945.08 \\
FTSE 100 & 2.0 & 64 & 16 & $9.17\times10^{-6}$ & $4.20\times10^{-4}$ & 1076.47 \\
KOSPI & 2.0 & 64 & 63 & 0 & $7.28\times10^{-12}$ & 963.81 \\
Nikkei 225 & 2.0 & 64 & 62 & 0 & $1.46\times10^{-11}$ & 709.91 \\
S\&P 500 & 2.0 & 64 & 10 & $3.94\times10^{-5}$ & $3.60\times10^{-4}$ & 1241.13 \\
\bottomrule
\end{tabular}
\tabnote{The table audits saved per-start likelihoods from the memory-restricted
ALM fits ($\gamma_-=\gamma_+$). ``Near-best'' is the number of successful
initial/refinement runs within 0.001 log-likelihood units of the best run. Best--2nd and
Best--5th are top-cluster gaps. Full range uses successful initial starts only and can be
large because the randomized grid intentionally includes poor starts. Convergence is
judged from repeated near-best solutions rather than from the full randomized range.}
\end{table}

The FIAPARCH comparison uses 24 candidate starts per asset, including warm, local-random,
and broad-random starts. Table~\ref{supp:tab:fiaparchwarm} reports the selected start and
the gain over the best prior candidate. No asset relies on a single isolated optimum.

\begin{table}[hbt!]
\centering
\caption{FIAPARCH warm-start and multi-start diagnostic.}
\label{supp:tab:fiaparchwarm}
\begin{tabular}{lrrrrl}
\toprule
Asset & Starts & Selected & Gain over best prior & Near-best & Flag \\
\midrule
S\&P 500 & 24 & 7 & 0.0000 & 16 & No \\
FTSE 100 & 24 & 9 & 7.0095 & 2 & No \\
DAX & 24 & 12 & 4.3588 & 4 & No \\
Nikkei 225 & 24 & 13 & 0.2389 & 3 & No \\
KOSPI & 24 & 5 & 0.0012 & 21 & No \\
Bitcoin & 24 & 8 & 0.0000 & 12 & No \\
\bottomrule
\end{tabular}

\tabnote{``Selected'' identifies the retained candidate among 24 starts. ``Gain over
best prior'' is its log-likelihood minus the best candidate before final refinement.
``Near-best'' counts solutions within 0.01 log-likelihood units of the selected optimum;
``Flag'' indicates whether the optimum is supported by only a single near-best solution.}
\end{table}

\section{Additional stability diagnostics}

\begin{figure}[hbt!]
\centering
\includegraphics[width=0.78\linewidth]{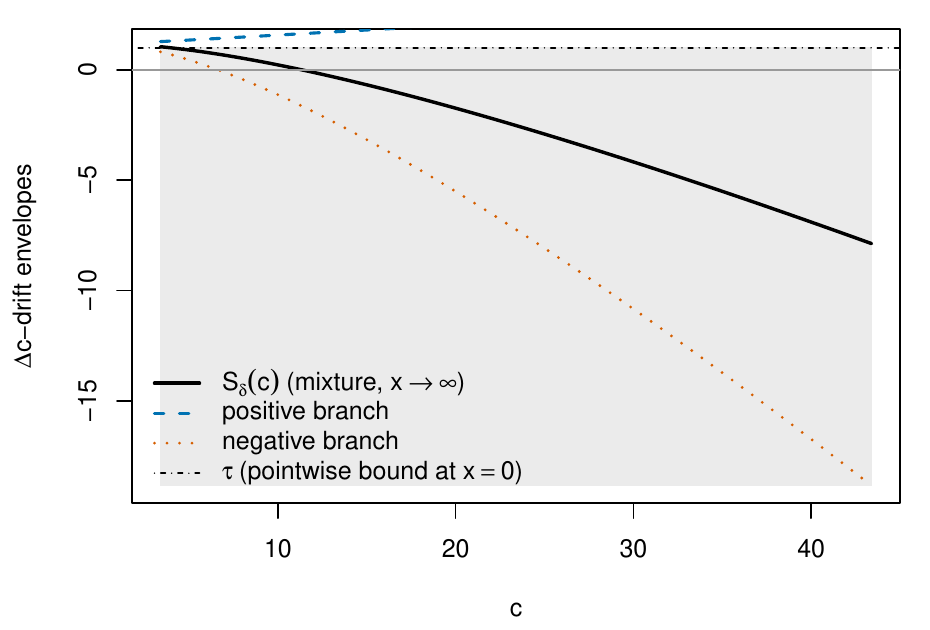}
\caption{Carrier-drift envelope at the fitted Nikkei~225 configuration.}
\label{supp:fig:cdrift}
\end{figure}

\begin{figure}[hbt!]
\centering
\includegraphics[width=\linewidth]{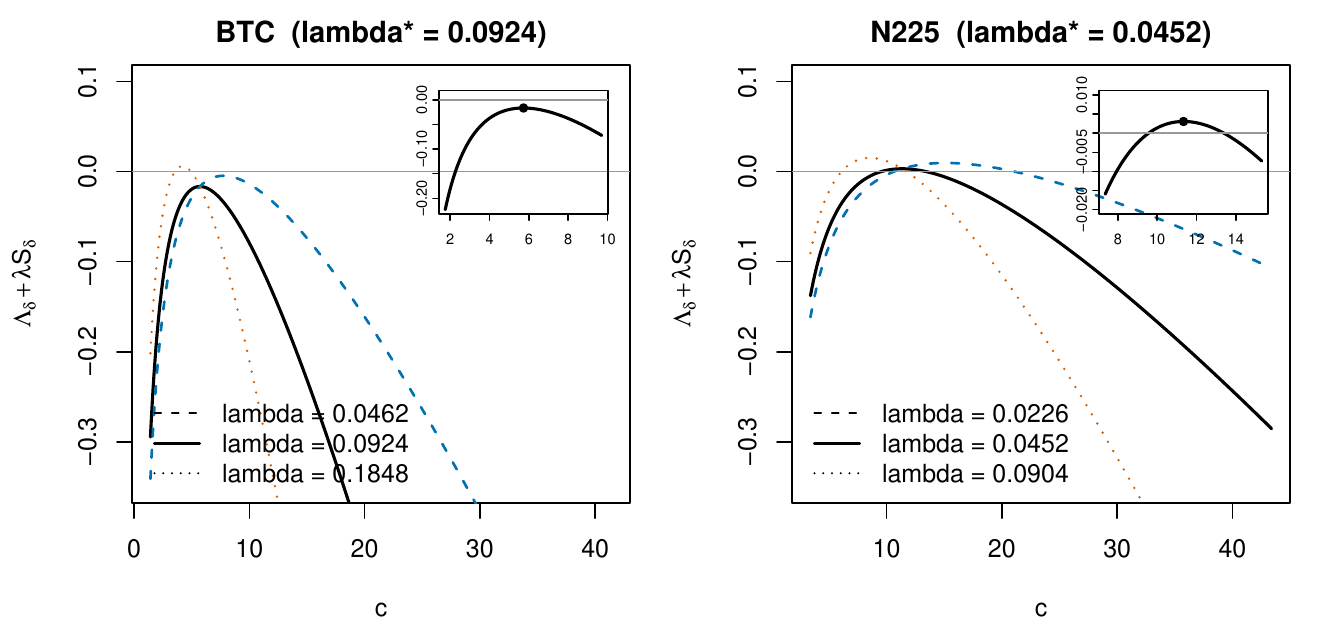}
\caption{Joint-stability diagnostic at the fitted Bitcoin and Nikkei~225
configurations.}
\label{supp:fig:joint}
\end{figure}

Figure~\ref{supp:fig:cdrift} shows the carrier-drift envelope for the fitted
Nikkei~225 configuration, and Figure~\ref{supp:fig:joint} compares the joint-stability
diagnostic for Bitcoin and the Nikkei~225.

\clearpage
\section{Full out-of-sample comparison}

A common-date check verifies finite-loss dates for all seven models under both QLIKE
and predictive likelihood. The numerical report and per-date forecast files are retained
in the replication materials so that the identical-date comparison in the main article
can be reproduced. Table~\ref{supp:tab:oosdm} reports the corresponding
Diebold--Mariano tests.

\begin{table}[H]
\centering
\caption{Diebold--Mariano tests against competing forecasts at fixed $p=1.2$.}
\label{supp:tab:oosdm}
\resizebox{\textwidth}{!}{%
\begin{tabular}{lrrrrrr}
\toprule
Asset & DM vs LM-GARCH & DM vs HAR-RV & DM vs GARCH & DM vs GJR & DM vs EGARCH & DM vs FIGARCH \\
\midrule
S\&P 500 & -0.82 & 5.41$^{***}$ & -1.10 & 1.38 & 3.05$^{***}$ & -0.14 \\
FTSE 100 & -0.62 & 0.62 & -0.58 & 0.13 & -0.92 & -0.24 \\
DAX & -2.03$^{**}$ & 2.90$^{***}$ & -2.11$^{**}$ & -3.60$^{***}$ & -0.07 & -1.75$^{*}$ \\
Nikkei 225 & -0.97 & 3.37$^{***}$ & -0.17 & -0.04 & 3.51$^{***}$ & 0.45 \\
KOSPI & -0.44 & 5.79$^{***}$ & -1.47 & -4.98$^{***}$ & -1.53 & 6.15$^{***}$ \\
Bitcoin & 1.17 & 2.58$^{***}$ & 1.01 & 0.44 & 1.58 & 4.81$^{***}$ \\
\bottomrule
\end{tabular}
}
\tabnote{The loss differential is $d_t=L_t^{\mathrm{ALM}}-L_t^{\mathrm{benchmark}}$,
so negative statistics favor ALM-GARCH. The long-run variance uses the
Newey--West/Bartlett estimator with lag $\lfloor n_{\mathrm{eval}}^{1/3}\rfloor$.
Stars denote two-sided significance: $^{*}p<0.10$, $^{**}p<0.05$, and
$^{***}p<0.01$.}
\end{table}

\end{document}